\pdfoutput=1
\documentclass[aps,
            10pt,
            prd,
            notitlepage,
            twocolumn,
            superscriptaddress,
            nofootinbib]{revtex4-2}

\usepackage{amsmath}
\usepackage{amssymb}
\usepackage{amsfonts}
\usepackage[utf8]{inputenc}
\usepackage[T1]{fontenc}
\usepackage{mathrsfs}
\usepackage{anyfontsize}

\usepackage[linktocpage,breaklinks]{hyperref}
\usepackage[usenames,dvipsnames]{xcolor}

\usepackage{newtxtext}
\usepackage{newtxmath}
\usepackage{tensor}
\usepackage{bm}

\usepackage{graphicx}
\usepackage{epsfig}
\usepackage{epstopdf}

\usepackage{natbib}
\usepackage{hyperref}

\hypersetup{colorlinks=true,
            citecolor=Plum,
            linkcolor=Plum,
            urlcolor=Plum}

\newcommand{\dd}{{\rm d}}

\newcommand{\pd}{\partial}

\newcommand{\RGB}{\mathcal{G}}
\newcommand{\order}[1]{\mathcal{O}(#1)}

\definecolor{EagleGreen}{RGB}{0,73,83}

\begin{document}
\title{The exterior spacetime of relativistic stars in scalar-Gauss-Bonnet gravity}

\begin{abstract}

The spacetime around compact objects is an excellent place to study gravity in the strong, nonlinear,
dynamical regime where solar system tests cannot account for the effects of large curvature. Understanding
the dynamics of this spacetime is important for testing theories of gravity and probing a regime which has
not yet been studied with observations. In this paper, we construct an analytical solution for the exterior
spacetime of a neutron star in scalar-Gauss-Bonnet gravity that is independent of the equation of state
chosen. The aim is to provide a metric that can be used to probe the strong-field regime near a neutron star
and create predictions that can be compared with future observations to place possible constraints on the
theory. In addition to constructing the metric, we examine a number of physical systems in order to see
what deviations exist between our spacetime and that of general relativity. We find these deviations to be
small and of higher post-Newtonian order than previous results using black hole solutions. The metric
derived here can be used to further the study of scalar-Gauss-Bonnet gravity in the strong field, and allow
for constraints on corrections to general relativity with future observations.
\end{abstract}

\author{Alexander Saffer}
\email{alexander.saffer@montana.edu}
\affiliation{eXtreme Gravity Institute,
Department of Physics, Montana State University, Bozeman, MT 59717 USA}

\author{Hector O. Silva}
\email{hector.okadadasilva@montana.edu}
\affiliation{eXtreme Gravity Institute,
Department of Physics, Montana State University, Bozeman, MT 59717 USA}

\author{Nicol\'as Yunes}
\email{nicolas.yunes@montana.edu}
\affiliation{eXtreme Gravity Institute,
Department of Physics, Montana State University, Bozeman, MT 59717 USA}

\date{\today}

\maketitle

\section{Introduction}
\label{sec:intro}

Einstein’s general relativity (GR) has proved to be an
exceptional theory to describe gravitational phenomena in
nature. From its early success in explaining the hitherto
mysterious advance of the perihelion of Mercury’s orbit
around the Sun~\cite{Will:2014kxa} to its consistency with the gravitational-wave
observations of merging black hole (BH) and neutron
star (NS) binaries by the LIGO/Virgo Collaboration~\cite{LIGOScientific:2018mvr},
GR has passed -- with flying colors -- all experiments it has
been confronted with.


Given the continual success of the theory, it is natural
to ask: should we consider GR as the final theory of the
gravitational interaction? Is it worth the effort to keep
developing further tests, seeking glimpses of a more
complete theory? Regarding the first question, field-theoretic
considerations have shown that GR is non-renormalizable,
placing a major obstacle to its quantization, and
indicating that the theory must be modified in the ultraviolet
regime. Indeed, a generic prediction of the low-energy
limits of quantum gravity theories, such as string
theory and loop quantum gravity, is that GR ought to be
augmented by both additional fields and higher-order
curvature scalars. Regarding the second question, GR’s
firm place in our vault of fundamental physical theories
implies that experimental evidence for a deviation would
shake the foundations of this vault.
Where should we search for signatures

Where should we search for signatures of beyond-GR
phenomenology? Compact objects, NSs and BHs, provide
a strong-field arena on which to put GR to the test in a
regime beyond the weak fields and low velocities of our
Solar System. The prototypical example is radio observations
of binary pulsars, which through the detailed and
careful monitoring of received pulses can reconstruct the
orbital motion of relativistic binaries to stupendous precision~\cite{Damour:2014tpa,Wex:2014nva,Kramer:2016kwa}. Another example of tests of GR with compact
objects is through the observation of electromagnetic
radiation emitted by the accretion disks that surround black
holes, although these tests are more challenging because of
the complex astrophysics in play during such observations
~\cite{Psaltis2008,Cardenas-Avendano:2019pec}. A final and more recent example is through the
observation of the x-ray pulse profile emitted by hot spots
on the surface of rapidly rotating stars~\cite{GendreauSPIE2012,ArzoumaninanSPIE2014,GendreauNature2017,Silva:2018yxz,Silva:2019leq}.
%
%

All of the tests mentioned probe the exterior spacetime of compact objects in one way or another.
Therefore, the construction of spacetimes close to these compact objects are required to place
constraints on theories that go beyond GR. While there are many modified theories that attempt
to explain anomalies between observations and the theoretical predictions of GR~\cite{Clifton:2011jh,Berti:2015itd},
a particularly interesting one is Einstein-dilaton-Gauss-Bonnet (EdGB) gravity. This theory is
interesting because it emerges in the low energy limit of heterotic string theory~\cite{Zhang:2017unx},
and it agrees well with GR in the weak field region~\cite{Sotiriou:2006pq}. EdGB gravity modifies the
Einstein-Hilbert action through the coupling of the Gauss-Bonnet invariant and a dynamical (dilaton)
scalar field~\cite{Mignemi:1992nt}. BH solutions in this theory have already been
developed~\cite{Kanti:1995vq,Torii:1996yi,Yunes:2011we,Ayzenberg:2014aka}, but until now,
NS solutions had only been obtained numerically~\cite{Pani:2011xm,Kleihaus:2014lba,Doneva:2017duq}.

In this paper, we present the first analytical solution of the field equations
in the small-dilaton expansion of EdGB gravity [i.e.~of scalar-Gauss-Bonnet
(sGB) gravity] that represents the exterior spacetime of non-rotating NSs,
working in the small coupling approximation. These solutions depend
\textit{only on the mass of the NS and the strength of the sGB coupling
parameter,} without any dependance on the dilaton scalar charge or any
additional constants of integration.  This is contrary to what has been found
in other theories containing a scalar field
(see.~\cite{Coquereaux:1990qs,Damour:1992we}), where a scalar charge depends on
integrals over the interior of the source.  The absence of this term allows our
final analytic solution to be implemented directly, without the need to
integrate the interior solution numerically to find the charge term.

With the known analytical exterior metric, we study the properties of this spacetime
by considering (timelike) geodesics, and derive sGB corrections to the innermost stable
circular orbit (ISCO), to the (circular) orbital frequency and to the epicyclical radial
frequencies of perturbed circular orbits. We also consider null geodesics and derive sGB
corrections to the visible fraction of a NS hot spot as observed from spatial infinity.

\begin{figure}[t]
    \includegraphics[width=\columnwidth,clip=true]{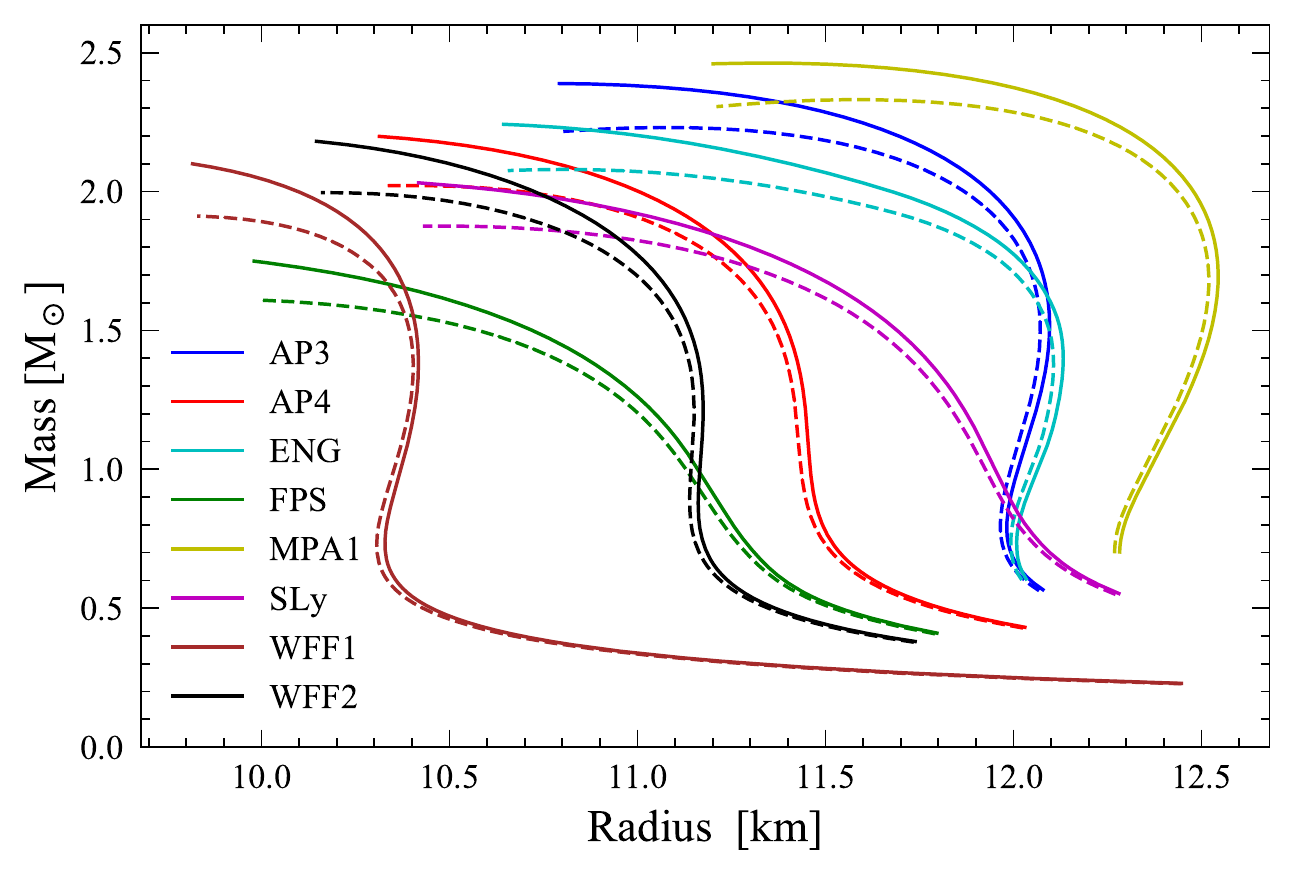}
    \caption{Mass-radius curves for various equations of state.
    The solid lines represent the GR solution, while the dashed lines correspond
    to our sGB solutions for $\alpha=15\kappa M_{\odot}^2$.  The central densities
    of the stars shown here range between $0.5$-$2.3 \times 10^{15} \,
    \rm{g}/cm^3$.}
\label{fig:MRCurves}
\end{figure}
A representative result is shown in Fig.~\ref{fig:MRCurves}, where we
present the sGB-corrected mass-radius relation of NS for different values
of the sGB coupling parameter and different equations of state (EoSs).
Each point in this plane represents a NS solution of a given total
gravitational mass and a given total radius, fixed through a numerical
integration of a given central density that requires the metric be
asymptotically flat at spatial infinity and $C^{1}$ everywhere.
Observe that the largest deviations arise in the high compactness regime
of the mass-radius relation, where the central densities are highest.
This makes sense given that sGB gravity introduces higher curvature
corrections to GR, which are bound to be largest when the compactness is
as large as possible.

The structure of this paper is as follows.
Section~\ref{sec:sGBintro} presents the basics of sGB gravity, including its
action and its field equations.
Section~\ref{sec:ExtSolution} explains our approach to finding a NS solution in
sGB gravity for the exterior spacetime, while Sec.~\ref{sec:IntSolution}
focuses on the interior regime and it presents the numerical solution to the
interior fields.
Section~\ref{sec:applications} discusses some astrophysical applications
that differ from the known results of GR.
Finally, Sec.~\ref{sec:conclusions} concludes and discusses how this
metric may be used in the future.
In the remainder of this paper, we use the $(-,+,+,+)$
metric signature and the conventions of~\cite{Misner:1974qy}, as
well as units in which $c=1=G$.

\section{Scalar Gauss-Bonnet gravity}
\label{sec:sGBintro}

In this section, we present the action of sGB gravity and its field equations.
We then introduce the perturbative scheme we will employ to analytically
solve the field equations, and we conclude by presenting the perturbatively
expanded field equations.

\subsection{Action}
We start by considering the action of a class of theories that contain
modifications proportional to the Gauss-Bonnet invariant, whose taxonomy
was described in~\cite{Yagi:2015oca}:
\begin{equation}
\label{eq:AllAction}
	S = S_{\rm EH} + S_{\varphi} + S_{\rm GB} + S_{\rm m}
\end{equation}
where $S_{\rm EH}$ is the Einstein-Hilbert action given by
\begin{equation}
S_{\rm EH} \equiv \kappa \int \dd^4x \sqrt{-g} \, R \,,\\
\label{eq:GRaction}
\end{equation}
with $\kappa \equiv (16\pi)^{-1}$, $g$ is the determinant of the metric $g_{ab}$,
$R \equiv g^{ab}R_{ab} = g^{ab}\tensor{R}{_a_c_b^c}$ is the Ricci scalar
(with $R_{ab}$ and $R_{abcd}$ being the Ricci and Riemann tensors respectively) and
\begin{equation}
S_{\varphi} \equiv -\frac{1}{2} \int \dd^4x \sqrt{-g} \, \left[\nabla_a \varphi \nabla^a \varphi + 2 U\left(\varphi\right)\right]\,.\\
\label{eq:Phiaction}
\end{equation}
is the action of a canonical scalar field $\varphi$ with potential $U$.
The coupling between the scalar field and the Gauss-Bonnet density
\begin{equation}
\RGB \equiv R^2 - 4 R_{ab}R^{ab} + R_{abcd}R^{abcd}
\label{eq:GaussBonnetTerm}
\end{equation}
is given by
\begin{equation}
S_{\rm GB} \equiv \int \dd^4x \sqrt{-g} \, \alpha \, f(\varphi)\,  \RGB\,,
\label{eq:QuadradicAction1}
\end{equation}
where $f(\varphi)$ specifies the functional form of the coupling and
$\alpha$ (with dimensions of [length]$^2$) its strength.
Finally, $S_{\textrm{m}}$ is the action of matter fields minimally coupled
to the metric.

The choice of $f(\varphi)$ defines the particular member in the class of Gauss-Bonnet theories.
For example, EdGB gravity is defined via $f(\phi)=e^{\phi}$, with typically a
massless dilaton so $U(\varphi) = 0$. Other coupling function $f(\phi)$ were also introduced
in~\cite{Antoniou:2017acq,Antoniou:2017hxj} and in the context of spontaneous black hole
scalarization in Refs.~\cite{Doneva:2017bvd,Silva:2017uqg,Doneva:2017duq,Silva:2018qhn}.
Hereafter, we expand $f(\varphi)$ in a Taylor series
$f(\varphi)=f(0) + f_{,\varphi}(0)\varphi + \order{\varphi^2}$ and
work in the so-called decoupling limit of the theory~\cite{Yagi:2015oca}.
Since $\RGB$ is a topological density, the first term in the series yields
a boundary term to the action which does not contribute to the equations
of motion. In the second term, $f_{,\varphi}(0)$ can be absorbed into the definition of
$\alpha$ and we obtain:
\begin{equation}
S_{\rm GB} \equiv \int \dd^4x \sqrt{-g} \, \alpha \, \varphi \, \RGB\,.
\label{eq:DEdGB}
\end{equation}

The action in Eq.~\eqref{eq:AllAction} with $S_{\rm GB}$ given by Eq.~\eqref{eq:DEdGB},
is sometimes called \textit{decoupled dynamical Gauss-Bonnet gravity}~\cite{Yagi:2015oca}
or more simply sGB gravity in this paper. This theory is invariant under constant shifts
$\varphi \to \varphi + c$ when $U = 0$~\cite{Sotiriou:2013qea,Sotiriou:2014pfa,Saravani:2019xwx},
and thus, it belongs to shift-symmetric Horndeski gravity~\cite{Kobayashi:2011nu}.
In this paper, we will restrict attention to sGB gravity with $U(\varphi) = 0$.

Here we here work in the \emph{small-coupling approximation} in which sGB modifications
are small relative to GR predictions. This approximation can be enforced by requiring that
$\alpha/\ell^2 \ll 1$, where $\ell$ is the characteristic length of our system.
For isolated NSs, the characteristic length scale is $\ell = \sqrt{R^3/M} = R/\sqrt{\mathscr{C}}$,
where $R$ is the radius of the star, $M$ its mass and $\mathscr{C}$ its compactness.
This length scale suggests the introduction of the dimensionless
coupling parameter\footnote{Note that this dimensionless coupling parameter is different
from that chosen in other work~\cite{Yagi:2012gp}, since here we normalize
$\alpha$ by $M_{\odot}$ instead of $M$.}
\begin{equation}
	\bar{\alpha} = \frac{\alpha}{\kappa\,M^2_\odot}\,,
\end{equation}
in terms of which the small coupling approximation reduces to
$\bar{\alpha} \ll \kappa^{-1} (R/M_{\odot})^{2} \mathscr{C}^{-1}$.
For NSs with $R \sim 11$ km and $\mathscr{C} \sim 0.2$, the small coupling
approximation then requires that $\alpha \ll 600 \; {\rm{km}}^{2}$ or
equivalently $\bar{\alpha} \ll 1.5 \times 10^{4}$.
The small-coupling approximation is well-justified because of current
constraints on $\alpha$. Observations of the orbital decay of
low-mass x-ray binaries~\cite{Yagi:2012gp} require that
$\alpha < 9 \; {\rm{km}}^{2}$, or $\bar{\alpha} < 220$.
When making plots, we will here work with $\bar{\alpha} \in (0,30)$, which
satisfies both the small coupling approximation and current constraints from
low-mass x-ray binary observations.

\subsection{Field equations}

We can obtain the field equations of the theory by varying the action $S$
with respect to the metric and the scalar field, with the result
\begin{subequations}
\begin{align}
\label{eq:FE1}
G_{ab}  &= - \frac{\alpha}{\kappa}\mathcal{K}_{ab} + \frac{1}{2\kappa}\left(T_{ab}^{\rm m} + T_{ab}^\varphi\right)\,, \\
\label{eq:scalar_field_eom}
\Box \varphi - U_{,\varphi} &= -\alpha \, \RGB\,,
\end{align}
\end{subequations}
where $G_{ab}$ is the Einstein tensor,
\begin{align}
\mathcal{K}_{ab} &= -2R \nabla_a\nabla_b \varphi +2\left(g_{ab}R-2R_{ab}\right) \Box \varphi + 8R_{c(a}\nabla^c \nabla_{b)}\varphi \nonumber \\
&\quad - 4g_{ab}R^{cd}\nabla_c \nabla_d\varphi + 4R_{acbd}\nabla^c \nabla^d \varphi\,,
\end{align}
while the stress-energy for the scalar field is
\begin{equation}
T^\varphi_{ab} = \nabla_a \varphi \nabla_b \varphi - \frac{1}{2}g_{ab}\left[\nabla_c \varphi \nabla^c \varphi - 2U(\varphi)\right]\,.
\end{equation}
As we stated before, we will here choose the scalar field to be massless and not
self-interacting, meaning that we can set $U = 0 = U_{,\varphi}$ in
the field equations.

Since we are interested in obtaining NS solutions in this theory,
we assume that matter is described by a perfect fluid, whose
stress-energy tensor is
\begin{equation}
\label{eq:MatterSET}
T^{ab}_{\rm m} = \left(\varepsilon + p \right)u^a u^b + p\,g^{ab}\,,
\end{equation}
where $u^{a}$ is the four-velocity of the fluid
(with pressure $p$ and total energy density $\varepsilon$)
subject to the constraint $u^{a} u_{a} = -1$.
Due to the diffeomorphism invariance of the theory, $T^{ab}_{\rm m}$
satisfies the conservation law
\begin{equation}
\nabla_a T^{ab}_{\rm m}=0\,,
\label{eq:EnergyConservation}
\end{equation}
as can be verified directly by taking the divergence of the field equations and using the equations of motion for the scalar field.

The EoS of cold nuclear matter characteristic of old NSs can be
well approximated by a barotropic EoS, that is $p = p(\varepsilon)$.
The large uncertainties on the properties of matter
in NS interiors result in a wide variety of competing EoS models~\cite{Lattimer:2015nhk}.
Here, to remain agnostic on which EoS correctly describes NS interiors
we consider eight different EoSs, which cover a wide range
of underlying nuclear physics models.
In increasing order of stiffness we use: FPS~\cite{Akmal:1998cf},
SLy~\cite{Douchin:2001sv}, WFF1~\cite{PhysRevC.38.1010}, WFF2~\cite{PhysRevC.38.1010},
AP4~\cite{Akmal:1998cf}, ENG~\cite{Engvik:1995gn}, AP3~\cite{Akmal:1998cf}, and
MPA1~\cite{Muther:1987xaa}.

\subsection{Perturbative expansion for the metric and fluid variables}
\label{sec:perturbative_scheme}

Having obtained the field equations, we now present the perturbative
scheme that we will use throughout this work. This approach was first introduced
in~\cite{Mignemi:1992nt} and was used in a number of studies involving
BHs~\cite{Yunes:2011we,Pani:2011gy,Sotiriou:2014pfa,Witek:2018dmd}.
Here, we apply this scheme for the first time to relativistic stars.

Let us consider a static, spherically symmetric star with
spacetime described by the line element
\begin{equation}
	\dd s^2 = -e^{2\tau}\dd t^2 + e^{2\sigma}\dd r^2 + r^2 \dd \Omega^2\,,
\end{equation}
where the metric functions $\tau$ and $\sigma$ contain only radial
dependence, and $\dd \Omega^{2} = \dd \theta^{2} + \sin^{2} \dd\phi^2$ is
the line element of the unit two-sphere.
The first step in the small-coupling approximation is to
expand all variables
$\vec{z} \in \{ \tau,\sigma,\varphi,\varepsilon,p \}$
in a power series in $\bar{\alpha}$ as follows
\begin{equation}
    \vec{z}(r) = \sum_{n = 0}^{N} \vec{z}_{n}(r)\,,
\end{equation}
where the subscript $n$ determines the power of $\bar{\alpha}$ associated with $\vec{z}_{n}$, i.e.
$\vec{z}_n = \order{\bar\alpha^n}$.

With these expansions, we can immediately make a few observations.
First, at ${\cal{O}}(\bar\alpha^0)$, the scalar field is everywhere constant,
because its source is zero [cf.,~Eq.~\eqref{eq:scalar_field_eom}].
We can then exploit shift-symmetry to impose $\varphi_0 = 0$. Second,
at ${\cal{O}}(\bar\alpha^1)$, we have $\tau_{1} = \sigma_{1} = 0$.
This follows from the fact that $\varphi_0 = 0$ and by Eq.~\eqref{eq:FE1},
the Einstein equations are identical to those of GR at this order. Furthermore,
since the metric is unaffected to this order and there is no direct coupling
between $\varphi$ and matter, we also have that $\varepsilon_1 = p_1 = 0$.

In this paper, we will obtain solutions for all variables $\vec{z}$ up to $\order{\bar\alpha^{2}}$.
From the proceeding discussion, we can outline the steps of the calculation ahead
as follows:
\begin{enumerate}
    \item at ${\cal{O}}(\bar\alpha^{0})$, the problem is identical to GR
        and we have to calculate $\{p_0, \varepsilon_0, \tau_0, \sigma_0\}$;
    \item at ${\cal{O}}(\bar\alpha^{1})$, we have to determine $\varphi_1$ on the background
        of a GR star obtained in the previous step;
    \item at ${\cal{O}}(\bar\alpha^{2})$, we must take into account the backreaction of
        the scalar field $\varphi_1$ onto the star to calculate
        $\vec{z}_{2} \in \{p_2, \varepsilon_2, \tau_2, \sigma_2\}$. The first two quantities
        tell us how the fluid is redistributed, while the latter how the spacetime
        is modified relative to the background GR metric.
\end{enumerate}

The perturbative scheme outlined above could be carried out to higher orders.
For instance, at $\order{\alpha^{3}}$ we would need to calculate $\varphi_{3}$ using the solutions
for $\vec{z}_{2}$.
Then, at $\order{\alpha^{4}}$, $\varphi_{3}$ would be used to obtain
$\vec{z}_{4}$.
We here stop our calculations at $\order{\bar\alpha^{2}}$ because this is the lowest order at which the metric
is modified, and therefore, the lowest-order we must have at hand if we want to calculate sGB corrections to astrophysical observables.

\subsection{Perturbative expansion of the field equations}

At $\order{\bar\alpha^0}$, the GR limit of the field equations give
\begin{equation}
G_{ab}^{0} = \frac{1}{2\kappa}\left[\left(\varepsilon_0 + p_0\right)u_au_b + p_0 g_{ab}\right]\,.
\label{eq:InsideStar}
\end{equation}
As usual~\cite{Wald:1984rg}, it is convenient to introduce a mass
function $m_{0} = (r/2)[1 - \exp(-2\sigma_0)]$, and then from the $(t,t)$ and
$(r,r)$-components of Eq.~\eqref{eq:InsideStar}, we find
\begin{subequations}
	\label{eq:TOVeqns}
	\begin{align}
		m_0' &= 4\pi \varepsilon_0 r^2\,,
        \label{eq:m_0}
        \\
		\tau_0'&= \frac{4 \pi p_0 r^3 + m_0}{r\left(r-2m_0\right)}\,,
        \label{eq:tau_0}
	\end{align}
\end{subequations}
Additionally, we can use the conservation law of Eq.~\eqref{eq:EnergyConservation}
to obtain
\begin{equation}
\label{eq:PressureEqn}
p_0' = \frac{\left(\varepsilon_0 +p_0\right)\left(4\pi p_0 r^3+m_0\right)}{r\left(2m_0-r\right)}\,,
\end{equation}
The system of equations~\eqref{eq:TOVeqns} and~\eqref{eq:PressureEqn}
are known as the Tolman-Oppenheimer-Volkoff (TOV) equations~\cite{Tolman:1939jz,Oppenheimer:1939ne}
and they are valid inside the star. The field equations outside the star can be
obtained from the set above through the limits $(\epsilon_{0},p_{0}) \to 0$.

At $\order{\bar\alpha^{1}}$, we have to solve the following equation
\begin{equation}
\label{eq:PhiFieldEqn}
\Box_{0} \varphi_1 = -\alpha \, \RGB_0\,,
\end{equation}
both inside and outside the star, where the d'Alembertian operator and the Gauss-Bonnet curvature invariant are
constructed from the metric functions found at $\order{\bar\alpha^0}$, i.e.
$\tau_0$ and $\sigma_0$.
Thus, equation~\eqref{eq:PhiFieldEqn} can be rewritten explicitly as
\begin{equation}
	\label{eq:NEWPhiFieldEqn}
	\frac{e^{-2\sigma_0}}{r}\left[\varphi_1''\,r + \varphi_1'\left(\tau_0'\,r-\sigma_0'\,r+2\right)\right] = -\alpha \RGB
\end{equation}

At $\order{\bar\alpha^2}$, the $(t,t)$ and $(r,r)$ components of the field equations yield
\begin{widetext}
\begin{subequations}
\label{eq:2ndOintFEs}
\begin{align}
\label{eq:Sig2FE}
&\left\{-64\pi\,\alpha\varphi_1'' + \left[64\pi\,\alpha\varphi_1' + 2\,r\left(2\tau_2 - 2\sigma_2 +1\right)\right]\sigma_0'+ 2\,r\sigma_2'-2\tau_2+2\sigma_2-1\right\}e^{-2\sigma_0} - 192\pi\,\alpha \varphi_1'\sigma_0'e^{-4\sigma_0} \nonumber \\
& \quad + 64\pi\,\alpha \varphi_1''e^{-4\sigma_0} + 2\tau_2+1 = 4\pi r^2\left[4\varepsilon_0 \tau_2 + \varphi_1'^2 e^{-2\sigma_0} + 2\left(\varepsilon_0 + \varepsilon_2\right)\right]\\
\label{eq:Tau2FE}
&-\left(1+2\sigma_2\right)e^{2\sigma_0} - 192\pi\,\alpha\varphi_1'\tau_0' e^{-2\sigma_0} + \left(64\pi \, \alpha \varphi_1' + 2r\right)\tau_0' + 2r\tau_2'+1 = 4\pi r^2\left[\left(4p_0 \sigma_2 + 2p_0 + 2p_2\right)e^{2\sigma_0} + \varphi_1'^2\right]\,,
\end{align}
\end{subequations}
\end{widetext}
while the conservation law of Eq.~\eqref{eq:EnergyConservation} gives
\begin{align}
\label{eq:ModTOV1}
p_2' &= -\frac{1}{r^2}\left[\left(\varphi_1''r^2 + \left(\tau_0' r^2-\sigma_0' r^2 + 2r\right)\varphi_1' \right. \right. \nonumber \\
&\left. \quad \left. -8\alpha\tau_0'' - 8\alpha \left(\tau_0'^2 - \tau_0'\sigma_0'\right)\right)\varphi_1'e^{-2\sigma_0} \right. \nonumber \\
&\left. \quad + 8\alpha\left(\tau_0'' + \tau_0'^2-3\tau_0'\sigma_0'\right)\varphi_1'e^{-4\sigma_0}\right] \nonumber \\
&\quad - \left(p_0 + p_2 + \varepsilon_0 + \varepsilon_2\right)\tau_0' + \left(p_0 + \varepsilon_0\right)\tau_2' + p_0'
\end{align}
in the stellar interior. The equations in the exterior
can be found through the limits $(\varepsilon_{0},\varepsilon_2,p_{0},p_2) \to 0$.

\section{Solutions of the field equations outside the star}
\label{sec:ExtSolution}

In this section, we first solve analytically, in vacuum, the equations presented
in Sec.~\ref{sec:sGBintro} order by order in $\bar\alpha$. The
general solutions to these equations will depend on integrations constants.
These constants can be fixed by examining the solutions' asymptotic behavior at
spatial infinity and imposing that (i) the spacetime is asymptotically flat
and that (ii) the scalar field approaches zero at spatial infinity.

\subsection{$\order{\bar\alpha^{0}}$ equations}

At this order, the solutions of Eqs.~\eqref{eq:m_0}--\eqref{eq:tau_0} have the usual
Schwarzschild form
\begin{equation}
    e^{2\tau_0} = e^{-2\sigma_0} = 1-\frac{a}{r} \,,
    \label{eq:metric_zeroth_order}
\end{equation}
where $a$ is an integration constant which (as we will see shortly)
is related with the gravitational mass $M$ of the star. In obtaining this solution,
we required that the metric be asymptotically flat near spatial infinity.

\subsection{$\order{\bar\alpha^1}$ equations}

At this order, we need to consider Eq.~\eqref{eq:PhiFieldEqn}. To
solve it, we first calculate $\RGB_0$ which can easily be found
using Eqs.~\eqref{eq:metric_zeroth_order} to be
\begin{equation}
\label{eq:RGBext}
\RGB_0 = \frac{12\,a^2}{r^6}\,,
\end{equation}
and in turn Eq.~\eqref{eq:NEWPhiFieldEqn} becomes
\begin{equation}
\label{eq:BoxPhi1}
r\left(a-r\right)\varphi_1'' + \left(a-2\,r\right)\varphi_1' =  \alpha\frac{12\, a^2}{r^4}\,,
\end{equation}
where Eq.~\eqref{eq:metric_zeroth_order} was used once again.

Equation~\eqref{eq:BoxPhi1} can be solved analytically to find
\begin{align}
\label{eq:PhiSol2}
\varphi_1 &=\frac{c_1}{a}\ln\left(1-\frac{a}{r}\right) + \frac{4\, \alpha}{a^2}\ln\left(1-\frac{a}{r}\right) \nonumber \\
&\quad + \frac{2\alpha}{r}\left(\frac{2}{a} + \frac{1}{r} + \frac{2a}{3r^2}\right) + c_2\,,
\end{align}
where $c_1$ and $c_2$ are two integration constants.
Requiring that the field vanishes at spatial infinity (i.e.~that the cosmological background value of the scalar
field is zero), we set $c_2 = 0$.
Expanding $\varphi_1$ about spatial infinity, we find that
\begin{equation}
\label{eq:phi_1_expansion}
\varphi_1 = - \frac{c_1}{r} - \frac{a\, c_1}{2\,r^2} - \frac{a^2\,c_1}{3\,r^3} + \order{r^{-4}}\,,
\end{equation}
which shows that $c_1$ is the scalar monopole charge.
Reference~\cite{Yagi:2015oca} showed that this charge vanishes for all stars,
and therefore, we can set $c_1 = 0$. The final expression for the scalar field outside the star is then
\allowdisplaybreaks[4]
\begin{equation}
\label{eq:phi_1_exterior}
\varphi_1 =\frac{4\, \alpha}{a^2}\ln\left(1-\frac{a}{r}\right) + \frac{2\,\alpha}{r}\left(\frac{2}{a} + \frac{1}{r} + \frac{2\,a}{3\,r^2}\right)\,.
\end{equation}

\subsection{$\order{\bar\alpha^2}$ equations}

At this order, we can substitute $\varphi_1$ [cf.~Eq.~\eqref{eq:phi_1_exterior}]
into Eqs.~\eqref{eq:2ndOintFEs}.
The resulting system of differential equations can be solved to find
\begin{subequations}
    \label{eq:metric_second_order}
	\begin{align}
    \label{eq:tau2ext}
	\tau_2 &= -\frac{3\zeta}{4}\left(1-\frac{7a}{6r}\right)\left(1-\frac{a}{r}\right)^{-1}\ln\left(1-\frac{a}{r}\right) \nonumber \\
           &\quad -d_1\,\zeta\,\frac{a}{2r} \left(1-\frac{a}{r}\right)^{-1} + d_2 \nonumber \\
	       &\quad -\zeta\,\frac{a}{r}\left(1-\frac{a}{r}\right)^{-1}\left(\frac{3}{4} - \frac{a}{2r}-\frac{3a^2}{16r^2}-\frac{5a^3}{48r^3} \right. \nonumber \\
           &\quad \left. -\frac{11a^4}{160\,r^4} - \frac{a^5}{20r^5} + \frac{5a^6}{48r^6}\right)\,,\\
    \label{eq:sig2ext}
	\sigma_2 &=  \frac{d_1 \, \zeta}{2}\left(\frac{a}{r}\right)\left(1-\frac{a}{r}\right)^{-1} - \frac{\zeta}{8} \, \frac{a}{r} \left(1-\frac{a}{r}\right)^{-1} \ln\left(1-\frac{a}{r}\right) \nonumber \\
	&\quad - \zeta \,\frac{a^2}{r^2}\left(1-\frac{a}{r}\right)^{-1}\left(\frac{1}{8} + \frac{a}{16r} + \frac{a^2}{24r^2} + \frac{a^3}{32r^3} \right. \nonumber \\
    &\quad \left. + \frac{a^4}{40r^4} - \frac{23a^5}{48r^5}\right)\,,
	\end{align}
\end{subequations}
where $d_1$ and $d_2$ are integration constants and we defined the dimensionless
parameter $\zeta$\footnote{Note with this definition of $\zeta$, the condition $\zeta \ll 1$ is not necessarily true.  The small coupling approximation requires that $\alpha/\ell^2 \ll 1$, and $\alpha/\ell^2 \neq \zeta^{1/2}$.}
via
\begin{align}
\label{eq:ZetaDef}
\zeta \equiv \frac{256 \pi  \alpha^2}{a^4}\,.
\end{align}\\

The constants of integration can be determined by studying the asymptotic behavior
of the metric functions about spatial infinity. For the $g_{tt}$ metric component we find
\begin{equation}
g_{tt} = e^{2 d_2} - \frac{a\left(1+d_1\,\zeta\right)}{r}\,e^{2 d_2} + \order{r^{-2}}\,,
\end{equation}
and thus, we set $d_{2} = 0$ without loss of generality, as any other choice corresponds
to a simple rescaling of the time coordinate $t \to t \exp(d_2)$.
From the $1/r$ term we identify
\begin{equation}
\label{eq:NewMass}
M \equiv \frac{a}{2}\left(1+d_1\,\zeta\right)\,.
\end{equation}
as a \textit{renormalized mass}: the gravitational
mass of the star that would be measured by an observer at spatial
infinity when performing a Keplerian observation.
Decomposing the mass via $M = M_0 + M_2$, we can identify $M_0 = a/2$
as the gravitational mass of a GR NS, and $M_2 = \zeta d_1 M_0$ as the sGB
correction to it. A similar mass renormalization occurs for black holes~\cite{Yunes:2011we}.

We can now reexpress our exterior solution in terms of the renormalized mass.
First, we eliminate $a$ in favor of $M$ in Eq.~\eqref{eq:NewMass} and
substitute the resulting equation into
Eq.~\eqref{eq:tau2ext}--\eqref{eq:sig2ext}. The resulting expressions for $\tau_2$ and
$\sigma_2$ can now be inserted in $g_{tt} = -\exp[2(\tau_0 + \tau_2)]$ and
$g_{rr} = \exp[2(\sigma_0 + \sigma_2)]$ and then, after
a reexpansion in powers of $\zeta$, we obtain our final expressions for
the metric up to $\order{\bar\alpha^{2}}$:
\allowdisplaybreaks[4]
\begin{widetext}
	\begin{subequations}
		\label{eq:FinalMetric}
\begin{align}
\label{eq:FinalMetric_gtt}
    g_{tt} &= -\left(1-\frac{2\,M}{r}\right) + \left[ \frac{3}{2}\left(1-\frac{7\,M}{3\,r}\right)\ln\left(1-\frac{2\,M}{r}\right) + \frac{M}{r}\left(3- \frac{4\,M}{r} - \frac{3\,M^2}{r^2} - \frac{10\,M^3}{3\,r^3} -\frac{22\,M^4}{5\,r^4} - \frac{32\,M^5}{5\,r^5} + \frac{80\,M^6}{3\,r^6}\right)\right] \zeta\,,\\
\label{eq:FinalMetric_grr}
    g_{rr} &= \left(1-\frac{2\,M}{r}\right)^{-1} -\frac{M}{r} \left(1-\frac{2\,M}{r}\right)^{-2}\left[ \frac{1}{2}\ln\left(1-\frac{2\,M}{r}\right) + \frac{M}{r} + \frac{M^2}{r^2} + \frac{4\,M^3}{3\,r^3} + \frac{2\,M^4}{r^4} + \frac{16\,M^5}{5\,r^5} - \frac{368\,M^6}{3\,r^6}\right]\zeta\,.
\end{align}
\end{subequations}
\end{widetext}
The equations~\eqref{eq:FinalMetric_gtt}--\eqref{eq:FinalMetric_grr} are independent of $d_1$, and are instead fully determined
by the mass $M$ of the star and the strength of the coupling constant
(through $\zeta$) only.

For consistency, let us now reexpress the scalar field also in terms of the renormalized mass.
The astute reader will notice that to $\order{\bar\alpha^{1}}$ we can simply replace
$a \to 2M$ in Eq.~\eqref{eq:phi_1_exterior} to obtain:
\begin{equation}
\varphi_1 =\frac{ \alpha}{M^2}\ln\left(1-\frac{2\,M}{r}\right)
+ \frac{2\,\alpha}{r}\left(\frac{1}{M} + \frac{1}{r} + \frac{4\,M}{3\,r^2}\right)\,,
\label{eq:phi_1_exterior_final}
\end{equation}
which is our final expression for the scalar field at $\order{\bar\alpha^{1}}$.
The sGB corrections to $a$ can be ignored in the scalar field, as they would
enter at $\order{\bar\alpha^{3}}$.

\subsection{Comparison with black holes spacetimes}
\label{subsec:com}

Before proceeding with the interior solution, let us compare the solutions obtained above
to their counterparts for BHs~\cite{Yunes:2011we}, focusing first on the scalar field solution.
The only difference between the calculation performed here and the one
carried out for BHs is that in the latter case $\varphi_1$
must be regular at the event horizon.
This results in a nonzero value of $c_1$ that yields
\begin{equation}
\varphi_1^{\rm BH}=\frac{2\,\alpha}{r}\left(\frac{1}{M_{\bullet}} + \frac{1}{r} + \frac{4\,M_{\bullet}}{3\,r^2}\right)\,,
\label{eq:phi_1_exterior_final_BH}
\end{equation}
which is identical to the second term in Eq.~\eqref{eq:phi_1_exterior_final}
with $M$ replaced by the hole's mass $M_{\bullet}$.
In a sense then, $\varphi_1$ is equal to $\varphi^{\rm BH}_1$ plus a
correction that arises from its continuity across the stellar surface.
We also observe that the BH limit of the NS solution for $\varphi_{1}$ is discontinuous.
One can see this easily by evaluating Eq.~\eqref{eq:phi_1_exterior_final} at the surface of
the star $R_0$ and taking the BH limit, $M/R \sim M_0 / R_0 \to 1/2$,
which is possible for certain anisotropic fluids in GR~\cite{Bowers:1974tgi,Raposo:2018rjn}.

Let us now compare the NS and BH exterior solutions for the exterior metric.
As in the case of the scalar field, requiring that the metric tensor be regular at the horizon
yields
\begin{subequations}
\label{eq:gtt_exterior_final_BH}
\begin{align}
g_{tt}^{\rm BH} = &-\left(1 - \frac{2M_{\bullet}}{r}\right)
\nonumber \\
&- \frac{1}{3} \frac{M_{\bullet}^{3}}{r^{3}} \left(1 + \frac{26 M_{\bullet}}{r}  + \frac{66}{5} \frac{M_{\bullet}^{2}}{r^{2}} + \frac{96}{5} \frac{M_{\bullet}^{3}}{r^{3}} - \frac{80 M_{\bullet}^{4}}{r^{4}} \right) \zeta\,,
\\
g_{rr}^{\rm BH} &= \left(1 - \frac{2 M_{\bullet}}{r}\right)^{-1} - \frac{M_{\bullet}^{2}}{r^{2}}  \left(1 - \frac{2 M_{\bullet}}{r}\right)^{-2}
\nonumber \\
&\times \left(1 + \frac{M_{\bullet}}{r} + \frac{52}{3} \frac{M_{\bullet}^{2}}{r^{2}} + \frac{2 M_{\bullet}^{3}}{r^{3}} + \frac{16}{5} \frac{M_{\bullet}^{4}}{r^{4}} - \frac{368}{3} \frac{M_{\bullet}^{5}}{r^{5}} \right) \zeta \,.
\end{align}
\end{subequations}
As in the case of the scalar field, the BH solution contains no logarithmic terms,
implying that the BH limit of the NS solution is singular. This is because of the
different choice of constants of integration in the NS and BH cases. Notice also that
in the BH case the metric differs from GR through terms of $\delta g_{tt}^{\rm BH} = {\cal{O}}(M_{\bullet}^{3}/r^{3})$
and $\delta g_{rr}^{\rm BH} = {\cal{O}}(M_{\bullet}^{2}/r^{2})$ in the far field, while in the
NS case it differs through terms of $\delta g_{tt} = {\cal{O}}(M^{7}/r^{7}) = \delta g_{rr}$.

\subsection{Note on the absence of a scalar charge in the NS solution}
In the case of other scalar-tensor theories, the work of~\cite{Coquereaux:1990qs,Damour:1992we}  shows that the exterior metric depends explicitly on a scalar charge, which in turn depends on the metric and matter terms within the star. If one wished to find the values of this charge, one would have to solve the TOV equations numerically for a specific set of initial conditions and a given equation of state. In our case, the exterior metric does not depend on any scalar charge, but rather it depends only on the mass, radius and coupling constant of the theory.
\section{Solutions of the field equations inside the star}
\label{sec:IntSolution}

For completeness, let us now tackle the problem of solving for the fluid variables, scalar field,
and metric components inside the star. This step will inevitably require
numerical integrations, for a relationship between pressure $p$ and
energy density $\varepsilon$ (i.e.~the EoS) must be given
and the resulting equations cannot be solved analytically.
In this section, we present the numerical scheme and the numerical solutions
for the interior fields. We stress however that
the exterior solutions found in the previous section do not require these interior
numerical solutions.

\subsection{$\order{\bar\alpha^{0}}$ equations}
As we saw in Sec.~\ref{sec:perturbative_scheme}, to this order we need to solve
the TOV equations of GR, i.e.~Eqs.~\eqref{eq:m_0}--\eqref{eq:tau_0} and~\eqref{eq:PressureEqn}.
We start by choosing an EoS from our catalog for which,
given a central total energy density $\varepsilon_{\rm c}$, gives
the corresponding central pressure $p_{\rm c} = p(\varepsilon_{\rm c})$.
We can then integrate Eqs.~\eqref{eq:m_0}--\eqref{eq:tau_0} and~\eqref{eq:PressureEqn} from
$r=0$ up to a point where $p_0(R_0) = 0$, which determines the star's radius
$R_0$.

In practice, we do this integration starting from a small, finite value of $r_{\rm c}$
and using a series solution valid in this region
\begin{subequations}
	\label{eq:InitialGRConditions}
	\begin{align}
	\label{eq:InitialGR1}
		m_0(r_{\rm c}) &= \frac{4\pi}{3}\,\varepsilon_c\,r_{\rm c}^3 + \order{r_{\rm c}^5}\,,\\
	\label{eq:InitialGR2}
		p_0(r_c) &= p_{\rm c} - \frac{2}{3}\left(3\pi \, p_{\rm c}^2 + 4\pi \,p_{\rm c} \,\varepsilon_{\rm c} \pi \, \varepsilon_{\rm c}^2\right)r_{\rm c}^2 + \order{r_c^4}\,,\\
	\label{eq:InitialGR3}
        \tau_0(r_{\rm c}) &= \tau_{0{\rm c}} + \left(2\pi \, p_{\rm c} + \frac{2\pi}{3}\,\varepsilon_{\rm c}\right)r_{\rm c}^2 + \order{r_c^4}\,.
	\end{align}
\end{subequations}
We terminate all integrations at the location where $p_0 / p_{\rm 0c} = 10^{-11}$.
The constant $\tau_{0{\rm c}}$ in the series solution of the metric is arbitrary
and is fixed \textit{a posteriori}.

At the star's surface $R_0$ we impose that the metric functions
$\tau_0$ and $\sigma_0$ are continuous, that is:
\begin{subequations}
\label{eq:metric_zeroth_bcs}
\begin{align}
    \label{eq:tau_match}
    \tau_0^{\rm in}(R_0) &= \tau_0^{\rm ext}(R_0)\,,\\
    \sigma_0^{\rm in}(R_0) &= \sigma_0^{\rm ext}(R_0)\,.
\end{align}
\end{subequations}
We can analytically match Eqs.~\eqref{eq:m_0} and~\eqref{eq:metric_zeroth_order}
at $R_0$ to find
\begin{equation}
\label{eq:aEQ2m}
	a=2m_0(R_0) \equiv 2 M_0\,,
\end{equation}
where $m_0(R_0)$ is the mass of the star enclosed inside the radius $R_0$.
Furthermore, Eq.~\eqref{eq:tau_match} fixes the value of the
constant $\tau_{\rm 0c}$. Our final numerical solution for $\tau_{0}$
corresponds to a simple shift $\tau_0 \to \tau_0 + \tau_{\rm 0 c}$.

The outcome of these integrations can be summarized in a mass-radius relation,
shown in Fig.~\ref{fig:MRCurves}. In this figure, the solid lines
correspond to various mass-radius curves for the EoSs in our catalog.

\subsection{$\order{\bar\alpha^{1}}$ equations}

At $\order{\bar\alpha^1}$ we only need to solve Eq.~\eqref{eq:PhiFieldEqn}.
From the $\order{\bar\alpha^{0}}$ solution, we know $a$ and $R_0$, which fully
determines $\varphi_1^{\rm ext}$ and its derivative at $R_0$
[cf.~Eq.~\eqref{eq:phi_1_exterior_final}].
This information can be used as initial conditions to integrate Eq.~\eqref{eq:PhiFieldEqn}
inside the star: we start our integration at $r=R_0$ and move in toward $r=0$.
In this calculation, it is useful to note that $\RGB_0$ is given by
\begin{equation}
    \RGB_0 = \frac{48\,m_0^2}{r^6} - \frac{128\pi(m + 2\pi r^3 p_0)}{r^6}\,,
    \label{eq:RGBint}
\end{equation}
inside the star~\cite{Silva:2017uqg}, where the functions $m_0$, $p_0$ and $\varepsilon_0$ are
all known from the $\order{\bar\alpha^{0}}$ calculation.

\begin{figure}[t]
\includegraphics[width=\columnwidth]{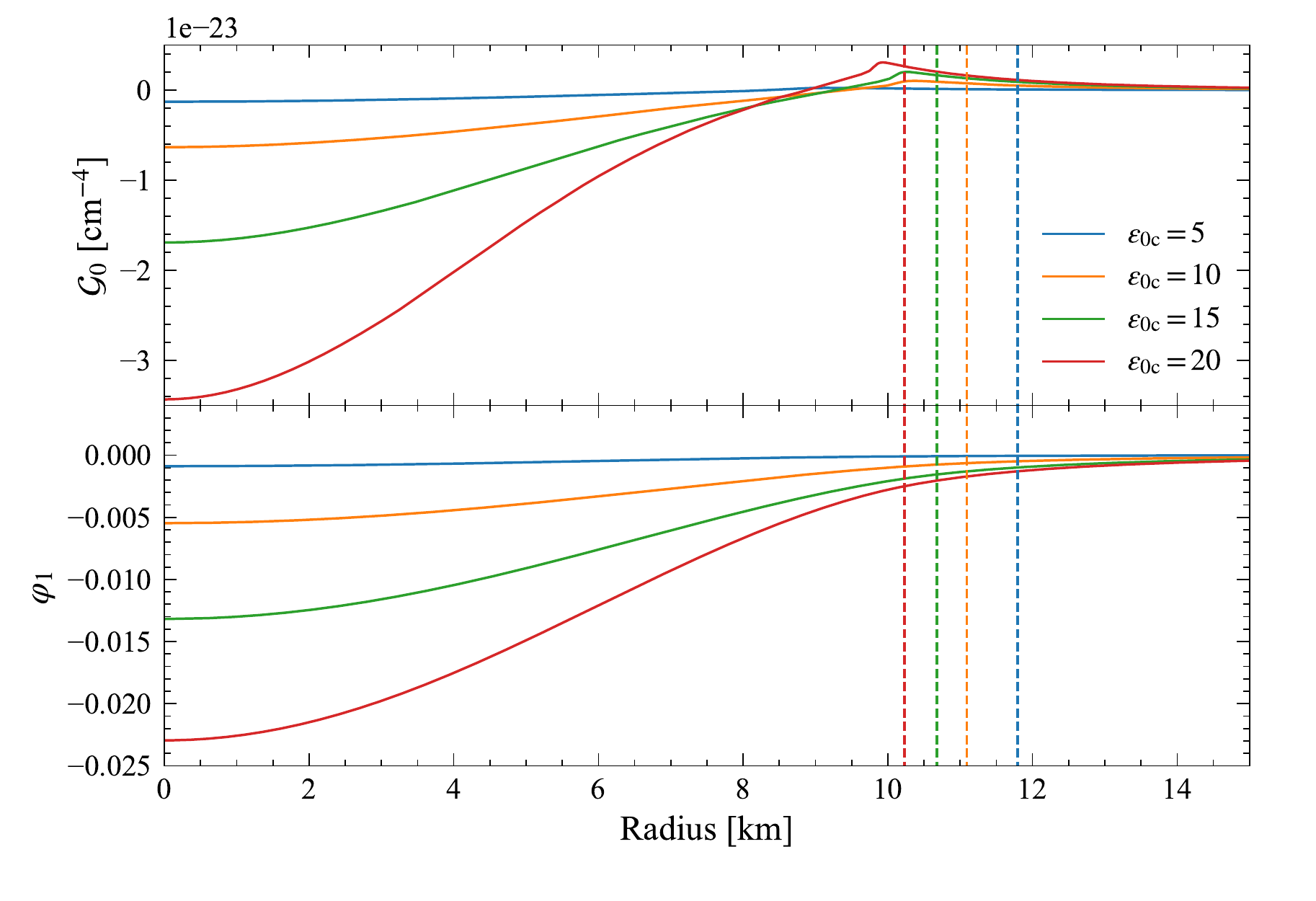}
\caption{Radial profiles of the Gauss-Bonnet invariant $\RGB_0$
(top) and the scalar field $\varphi_1$ (bottom) at $\order{\bar\alpha^{1}}$ for an SLy EoS. In both panels,
the different colors correspond to different central energy densities
$\varepsilon_{\rm 0 c}$ (in units of $10^{14} \, \text{g/cm}^3$).
The vertical dashed lines correspond to the radius for each star.
All scalar field solutions were calculated at a fixed $\bar{\alpha}=15$
coupling constant strength.}
\label{fig:FirstOrderPlots}
\end{figure}

The radial profiles of $\RGB_0$ and $\varphi_1$ are shown in
Fig.~\ref{fig:FirstOrderPlots} using the SLy EoS with the scalar-Gauss-Bonnet coupling
fixed to $\bar{\alpha}=15$. In the top-panel, we see
that $\RGB_0$ is mostly negative within the star, except near the surface (indicated by
the dashed vertical line) where it changes sign and then matches smoothly
to its exterior form, given in Eq.~\eqref{eq:RGBext}. We also observe that
$\RGB_0$ has a larger magnitude for stars with larger values of $\varepsilon_{\rm 0c}$.
This is can be seen by substituting the expansions of Eqs.~\eqref{eq:InitialGR1}--\eqref{eq:InitialGR3} into Eq.~\eqref{eq:RGBint}.
We find that $\RGB_0$ is negative and nearly constant close to the center of the star at $r \approx 0$,
with its magnitude proportional to $\varepsilon_{\rm 0 c}$.
In the bottom-panel, we see that NSs with larger central energy densities $\varepsilon_{\rm 0c}$ have
larger amplitudes of $\varphi_{1}$ at their cores. This is unsurprising given the fact that the
source of the scalar, i.e. $\RGB_0$, has a larger magnitude near the stellar center.
At the surface, $\varphi_1$ connects smoothly with its exterior solution, given by Eq.~\eqref{eq:PhiSol2}
(at this order in $\alpha$).
The results for other EoSs are qualitatively the same as the ones shown here.

\begin{figure}[t]
\includegraphics[width=\columnwidth]{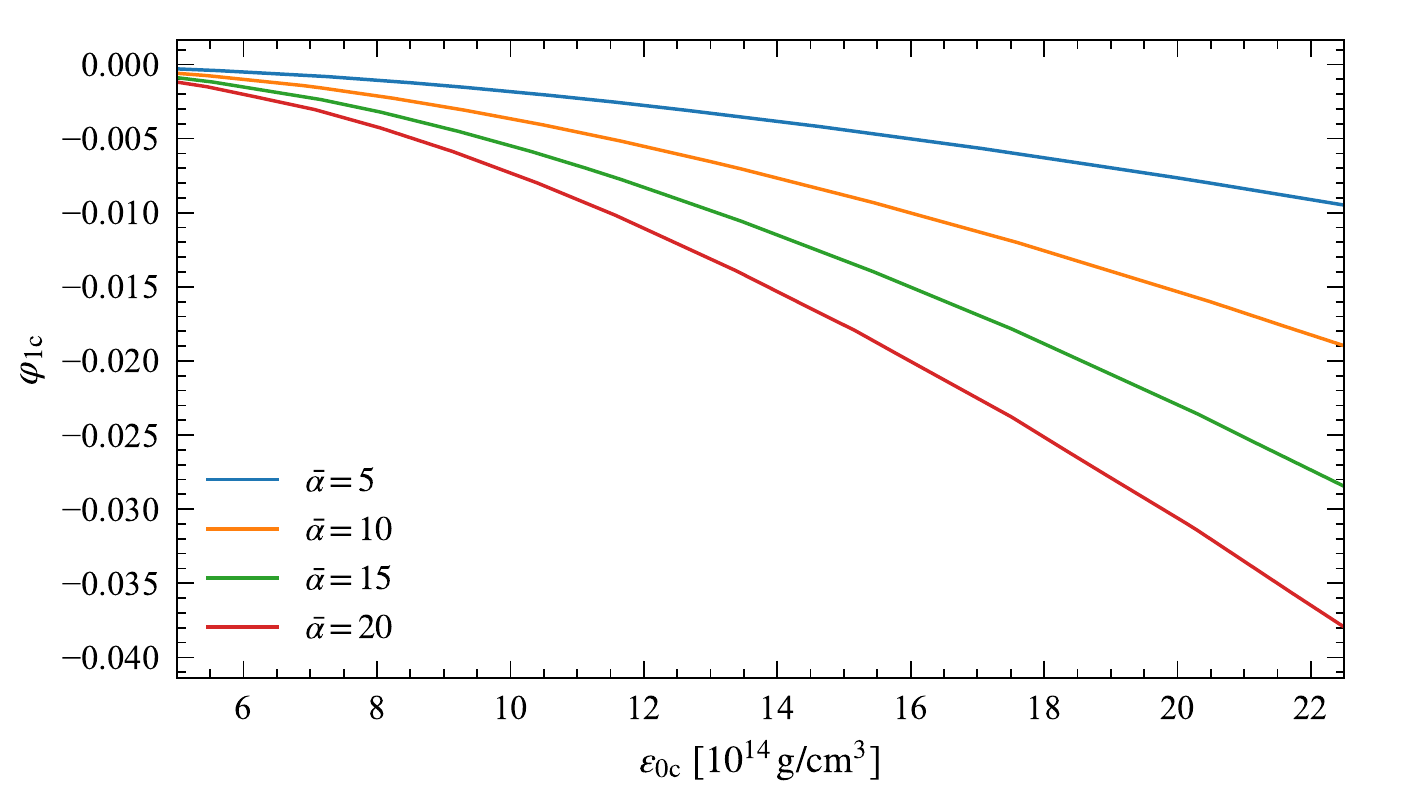}
\caption{Central values of the scalar field
$\varphi_1$ for various values of $\bar{\alpha}$ as a function of the
central densities of the star with an SLy EoS.  Observe how the central value
of the scalar field converges toward zero at small central densities irrespective
of the coupling constant.
}
\label{fig:CentralPhi}
\end{figure}

Let us now investigate how the central values of the scalar field $\varphi_1$
vary as a function of both $\varepsilon_{\rm 0c}$ and of $\bar{\alpha}$. This dependence
is shown in Fig.~\ref{fig:CentralPhi} for four representative values of $\bar{\alpha} = \{5,10,15,20 \}$
covering a range of central energy densities $\varepsilon_{\rm 0c}$ that span
stars with masses $0.552\,M_\odot$ to $2.03\,M_\odot$ using the SLy EoS.
We see that for small $\varepsilon_{\rm 0c}$ (i.e.~low-mass stars) all values of $\varphi_{\rm 1c}$
converge towards zero regardless of the strength of the coupling. This is can understood by noticing
that in this limit $\RGB_0$ is very small and nearly flat (cf.~Fig.~\ref{fig:FirstOrderPlots}),
thus sourcing $\varphi_1$ weakly. For larger $\varepsilon_{\rm 0c}$, the situation is
different and we see a stronger dependence of the central value of $\varphi_{1}$ on $\bar{\alpha}$.
Unsurprisingly, the magnitude of $\varphi_1$ is larger at the stellar core the larger the strength
of the coupling $\bar{\alpha}$.

\subsection{$\order{\bar\alpha^{2}}$ equations}

At $\order{\bar\alpha^2}$ we need to solve Eqs.~\eqref{eq:Sig2FE}--\eqref{eq:Tau2FE} and~\eqref{eq:ModTOV1}.
The boundary conditions are similar to those at $\order{\bar\alpha^0}$.
Imposing continuity at the surface gives us
\begin{equation}
	\label{eq:2ndOBC}
    g_{ab}^{\text{in}}(\alpha^0,\alpha^2,R_2) = g_{ab}^{\rm ext}(\alpha^0,\alpha^2,R_2)\,,
\end{equation}
where $R_2$ is the radius of the NS at $\order{\bar\alpha^2}$ given by the condition:
\begin{equation}
	\label{eq:PressureCondition2}
	p_0(R_2)+p_2(R_2) = 0\,.
\end{equation}
As in the $\order{\bar\alpha^{0}}$ integrations, we start from $r_{\rm c}$
and integrate outwards until the point $R_2$ where the condition
$(p_{2} + p_{0}) / p_{\rm 0 c} = 10^{-11}$ is met.
Equation~\eqref{eq:2ndOBC} allow us to determine the numerical value of $d_1$, which
in turn allows us to calculate the renormalized mass $M$ [Eq.~\eqref{eq:NewMass}] and
thereby determine the exterior metric in terms of interior quantities.

When integrating Eqs.~\eqref{eq:Sig2FE}--\eqref{eq:Tau2FE} and~\eqref{eq:ModTOV1} we need to
be careful on how we calculate the perturbed density $\varepsilon_2$.
To do this, we take our total density $\varepsilon(p) = \varepsilon_0 + \varepsilon_2$, and solve for $\varepsilon_2$ as
\begin{equation}
\label{eq:2ndOrderDensity}
	\varepsilon_2 = \varepsilon\left(p_0 + p_2\right) - \varepsilon_0\left(p_0\right)\,,
\end{equation}
where $\varepsilon(p_0 + p_2)$ is a spline interpolation of our EoS table.
This allows us to eliminate the variable $\varepsilon_2$ in favor of the
perturbed pressure $p_2$.
With Eq.~\eqref{eq:2ndOrderDensity} and the solutions to all fields
up to $\order{\bar\alpha}$, we can solve our system of equations
given by Eqs.~\eqref{eq:Sig2FE}--\eqref{eq:Tau2FE} and~\eqref{eq:ModTOV1}.

\begin{figure}[t]
	\includegraphics[width=\columnwidth]{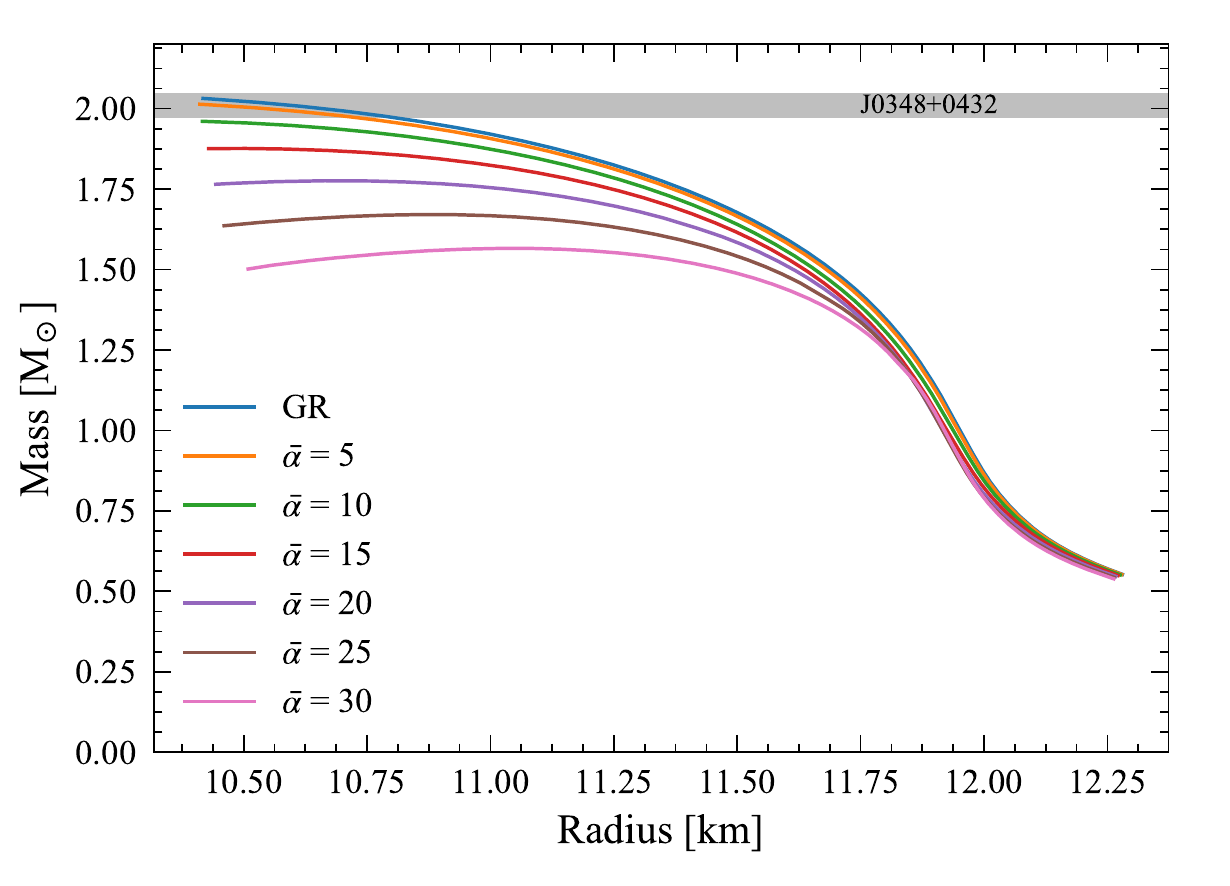}
    \caption{Mass-radius curves with a SLy EoS for varying couplings $\bar{\alpha}$.
    Observe that greater couplings lead to a decrease in the maximum mass of NSs, which can aid
    in constraining the theory with observations of massive pulsars.}
	\label{fig:MRalphaVary}
\end{figure}

The dashed curves in Fig.~\ref{fig:MRCurves} show the mass-radius relations calculated
to $\order{\bar\alpha^2}$ for the various EoSs of our catalog. We see that for a fixed value of
$\bar{\alpha}$ the deviations from the GR mass-radius relation
occur at larger masses. This is consistent with our previous observations on $\varphi_1$,
which had a larger magnitude for larger masses. Consequently, these large scalar fields backreact
more strongly onto the GR solution, causing larger changes to the mass and the radius.
The sGB corrections typically lead to less massive NSs regardless of the EoS considered,
a result consistent with those of~\cite{Pani:2011xm}.
For clarity, in Fig.~\ref{fig:MRCurves} we only showed curves with a fixed $\bar{\alpha} =15$,
but how do the mass-radius curves change (for a fixed EoS) as we vary $\bar{\alpha}$?
This is shown in Fig.~\ref{fig:MRalphaVary} for the SLy EoS. As expected, from our previous discussion
of the $\order{\bar\alpha^{1}}$ results, an increase in $\bar{\alpha}$ causes larger
deviations in the mass-radius curve. Indeed, the larger the value of the sGB coupling,
the smaller the maximum NS mass that is allowed for a given EoS.

Figure~\ref{fig:MRCurves} also shows vividly the difficulties of testing modified
theories of gravity with masses and radii measurements of NSs. In the absence of a complete
understanding of matter in the NS interior, the various competing EoS models
predict NSs that cover a wide portion of the $(M,R)$ plane.  But this problem
could be averted if, in the future, the EoS is constrained through
NICER~\cite{GendreauSPIE2012,ArzoumaninanSPIE2014,GendreauNature2017} and/or LIGO/VIRGO~\cite{TheLIGOScientific:2017qsa}.
Let us imagine, for example,
that the SLy EoS is favored by observations. If so, the observation
of a $\approx 2 M_{\odot}$ NS (see e.g.~\cite{Antoniadis:2013pzd}) would place the stringent constraint
$\bar{\alpha} \lesssim 10$ (roughly one order of magnitude more stringent than current bounds),
since for larger values a SLy EoS could not predict such a massive NS (see Fig.~\ref{fig:MRalphaVary}).
This constraint would be weaker if the true EoS is stiffer
(e.g.~MPA1 and AP3), as larger values of $\bar{\alpha}$ would be required to pull the
mass-radius curve below $\approx 2M_{\odot}$, but stiffer EoSs are disfavored
by recent tidal deformability constraints from the GW170817 gravitational-wave
event~\cite{Abbott:2018exr}.

Our results shown above are in agreement with those obtained previously in the
literature. For example, NS solutions in the full EdGB theory were presented
first in Ref.~\cite{Pani:2011xm}. We compared our final results
with those from Ref.~\cite{Pani:2011xm} (see e.g. Figs. 1-3 in that paper) and found
good agreement when the coupling was small (i.e. in the decoupling limit).
%

\section{Astrophysical applications}
\label{sec:applications}

Now that we have a analytic solution for the exterior spacetime of a NS in sGB gravity [see Eqs.~\eqref{eq:FinalMetric_gtt}--\eqref{eq:FinalMetric_grr}],
let us explore some astrophysical applications to investigate the physical effects of the corrections
on observables.

Probing astrophysical phenomena in the vicinity of NSs naturally requires that one
first analyze the geodesic motion of massive test particles and of light in the stellar exterior.
Since our metric is static and axisymmetric, we know it possesses
a timelike and azimuthal Killing vector, which imply the existence of two
conserved quantities: the specific energy $E$ and the specific angular momentum $L$
\begin{equation}
E = -g_{tt}\dot{t}\,, \quad
L = g_{\phi\phi}\dot{\phi}\,.
\label{eq:conserved_quants}
\end{equation}
where the dots indicate differentiation with respect
to proper time. From normalization condition of the four-velocity,
$u^{a} u_{a} = \epsilon$, where $\epsilon=(-1 \text{ or } 0)$ for time-like or null trajectories respectively, we obtain
\begin{equation}
\label{eq:rDotEqn}
	\frac{\dot{r}^2}{2} = V_{\rm eff}(r)\,,
\end{equation}
which describes the radial motion of the particle in terms
of the effective potential
\begin{equation}
\label{eq:VeffEqn}
	V_{\rm eff}(r) = -\frac{1}{2g_{rr}}\left(\frac{E^2}{g_{tt}} + \frac{L^2}{g_{{\phi} {\phi}}} - \epsilon\right)\,.
\end{equation}
Because of spherical symmetry, we can set $\theta = \pi / 2$ (and therefore
$g_{\phi\phi} = r^2$) without loss of generality.

\subsection{Circular orbits around the star}
Let us study the circular motion of massive test particles ($\epsilon=-1$) around a NS with exterior metric
given by Eqs.~\eqref{eq:FinalMetric_gtt}--\eqref{eq:FinalMetric_grr}.
For a circular orbit at $r=r_{\ast}$, the conditions $V_{\rm eff}(r_{\ast})=0$
and $V_{\rm eff}'(r_{\ast})=0$ must be satisfied.
Using Eq.~\eqref{eq:VeffEqn}, we can solve for $E$ and $L$, and
expand in powers of $\zeta$ to obtain
\begin{subequations}
	\label{eq:EandLexpansions}
	\begin{align}
	\label{eq:E_expand}
        E &= E_0 + \zeta  E_{2} + \order{\zeta^2}\,,\\
    \label{eq:L_expand}
        L &= L_0 + \zeta  L_{2} + \order{\zeta^2}\,,
	\end{align}
\end{subequations}
where $E_0$ and $L_0$ are the GR specific energy and angular momentum
for circular orbits~\cite{Wald:1984rg}
\begin{subequations}
    \label{eq:e_and_l_gr}
	\begin{align}
        E_0 &= \left(1-\frac{2M}{r_{\ast}}\right)\left(1-\frac{3\,M}{r_{\ast}}\right)^{-1/2}\,,\\
        L_0 &= \left(Mr_{\ast}\right)^{1/2}\left(1-\frac{2M}{r_{\ast}}\right)^{-1}E_0\,,
	\end{align}
\end{subequations}
and $E_{2}$ and $L_{2}$ are modifications of $\order{\bar\alpha^{2}}$.
The latter are given by
	\begin{align}
        E_2 &= - \left(1-\frac{3\,M}{r_{\ast}}\right)^{-3/2}\left[\frac{3}{4} - \frac{31\,M}{8\,r_{\ast}} + \frac{21\,M^2}{4\,r_{\ast}^2}\right]\ln\left(1-\frac{2\,M}{r_{\ast}}\right)\nonumber \\
	    & \quad -\frac{M}{r_{\ast}} \left(1-\frac{3\,M}{r_{\ast}}\right)^{-3/2}\left(\frac{3}{2}-\frac{25\,M}{4\,r_{\ast}}+\frac{19\,M^2}{4\,r_{\ast}^2} \right. \nonumber \\
        & \quad \left. + \frac{19\,M^3}{6\,r_{\ast}^3} + \frac{33\,M^4}{10\,r_{\ast}^4} + \frac{21\,M^5}{5\,r_{\ast}^5} -\frac{596\,M^6}{15\,r_{\ast}^6}
        + \frac{40\,M^7}{r_{\ast}^8}\right)\,,
     \end{align}
and
    \begin{align}
	L_2 &= \frac{\left(M\,r_{\ast}\right)^{1/2}}{8}\left(1-\frac{3\,M}{r_{\ast}}\right)^{-3/2}\ln\left(1-\frac{2\,M}{r_{\ast}}\right) \nonumber \\
        & \quad + \left(M\,r_{\ast}\right)^{1/2}\left(1-\frac{3\,M}{r_{\ast}}\right)^{-3/2} \left(\frac{M}{4\,r_{\ast}} + \frac{M^2}{4\,r_{\ast}^2}+\frac{M^3}{3\,r_{\ast}^3}\right. \nonumber \\
        & \left. \quad + \frac{M^4}{2\,r_{\ast}^4} + \frac{4\,M^5}{5\,r_{\ast}^5} - \frac{188\,M^6}{3\,r_{\ast}^6} + \frac{80\,M^7}{r_{\ast}^7}\right)\,.
	\end{align}

We may now make use of Eqs.~\eqref{eq:rDotEqn} and~\eqref{eq:VeffEqn} along
with our circular orbit conditions to find the sGB modifications to the location of the ISCO.
Doing so, we find that the ISCO radius is
\begin{equation}
	R_{\rm ISCO} = 6M - \frac{3M}{2}\left[\frac{5047}{14580} + \ln\left(\frac{2}{3}\right)\right]\zeta + \order{\zeta^2}\,.
\end{equation}
which reproduces the well-known GR result when $\zeta = 0$. Notice that the sGB correction
pushes the ISCO location farther away from the stellar surface (assuming the star is sufficiently
compact so that the ISCO is outside the surface in GR in the first place) by a
small amount $R_{\rm ISCO} - 6 M \approx + 0.089 M \zeta$.

\subsection{Modified Kepler's third law}
Now let us derive an expression for the orbital frequency
$\Omega_{\phi} = \dd \phi / \dd t$ of a massive particle in circular orbit at radius $r_{\ast}$
as measured by an observer at infinity.
Using Eqs.~\eqref{eq:conserved_quants} and~\eqref{eq:E_expand}--\eqref{eq:L_expand} we find
 	\begin{align}
 	\label{eq:Kepler}
     \Omega_{\phi}^2/\Omega_{0}^2 - 1 =& - \frac{7}{4}\ln\left(1-\frac{2\,M}{r_{\ast}}\right) \zeta - \left(1-\frac{2\,M}{r_{\ast}}\right)^{-1}
     \nonumber \\
     &\times
     \left(\frac{7\,M}{2\,r_{\ast}} - \frac{7\,M^2}{2\,r_{\ast}^2} -\frac{7\,M^3}{3\,r_{\ast}^3}-\frac{7\,M^4}{3\,r_{\ast}^4} - \frac{14\,M^5}{5\,r_{\ast}^5}
     \right.
     \nonumber \\
     &\left.
      - \frac{1976\,M^6}{15\,r_{\ast}^6} + \frac{560\,M^7}{3\,r_{\ast}^7}\right) \zeta + \order{\zeta^2}\,,
 	\end{align}
where $\Omega_{0}^2 = M/r_{\ast}^3$ is the usual GR result.
Expanding Eq.~\eqref{eq:Kepler} in the far field limit we find to leading order in $\zeta$
\begin{equation}
\label{eq:KeplerApprox}
	\Omega_{\phi}^2 \approx \Omega_{0}^2 \left(1+\frac{128M^6}{r_{\ast}^6}\,\zeta \right)\,,
\end{equation}
which is consistent with our expansions of the sGB metric deformation in Sec.~\ref{subsec:com}.
Unlike the case for BHs, where the correction to the frequency occurs as
$\order{M^2/r^2}$~\cite{Yunes:2011we}, the presence of the logarithmic term in
Eq.~\eqref{eq:Kepler} gives a small correction. This suggests that weak-field observables
will be very poor probes of sGB gravity.

\subsection{Quasiperiodic oscillations}

Let us now focus on the frequencies of quasiperiodic oscillations (QPOs).
There are a number of models which have been proposed as possible causes of
QPOs including the relativistic motion of matter~\cite{Stella:1998mq} and
resonance between orbital and epicyclic motion~\cite{Abramowicz:2001bi}.  Regardless of
the model in question, it may be interesting to calculate the sGB corrections the QPO frequencies
to study what effect, if any, this modification to GR has.

The orbital frequency was already calculated in Eq.~\eqref{eq:KeplerApprox}, so let us now
calculate the epicyclic frequency for timelike geodesics. This frequency is
determined by a radially-perturbation to the circular orbit equation [Eq.~\eqref{eq:rDotEqn}],
which yields
\begin{equation}
\label{eq:EpyDef}
	\Omega_r^2 = -\frac{1}{2\,\dot{t}}\frac{\pd^2 V_{\rm eff}(r)}{\pd r^2}\,.
\end{equation}
Solving Eq.~\eqref{eq:EpyDef} with the $V_{\rm eff}(r)$ defined in Eq.~\eqref{eq:VeffEqn} gives us
\allowdisplaybreaks[4]
	\begin{align}
		\Omega_r^2/\Omega_{0}^{2} &= 1-\frac{6\,M}{r} - \frac{7}{4}\left(1-\frac{48\,M}{7\,r}\right)\ln\left(1-\frac{2\,M}{r}\right) \zeta
		\nonumber \\
		&\quad+\frac{M}{r} \left(1-\frac{2\,M}{r}\right)^{-1}\left(\frac{7}{2} - \frac{55\,M}{2\,r} + \frac{65\,M^2}{3\,r^2} + \frac{41\,M^3}{3\,r^3}
		\right.
		\nonumber \\
		&\quad\left.
		+ \frac{66\,M^4}{5\,r^4} + \frac{9832\,M^5}{15\,r^5} - \frac{15056\,M^6}{15\,r^6} - \frac{256\,M^7}{r^7}\right) \zeta
        \nonumber \\
        &\quad+ \order{\zeta^2}\,.
	\end{align}
If one were to asymptotically expand this frequency about spatial infinity,
one would again find that the sGB corrections are highly suppressed. As we will see below,
however, QPOs are sensitive to physics near the ISCO, and in this regime, the sGB
corrections are not nearly as suppressed.

\begin{figure}[t]
	\includegraphics[width=\columnwidth]{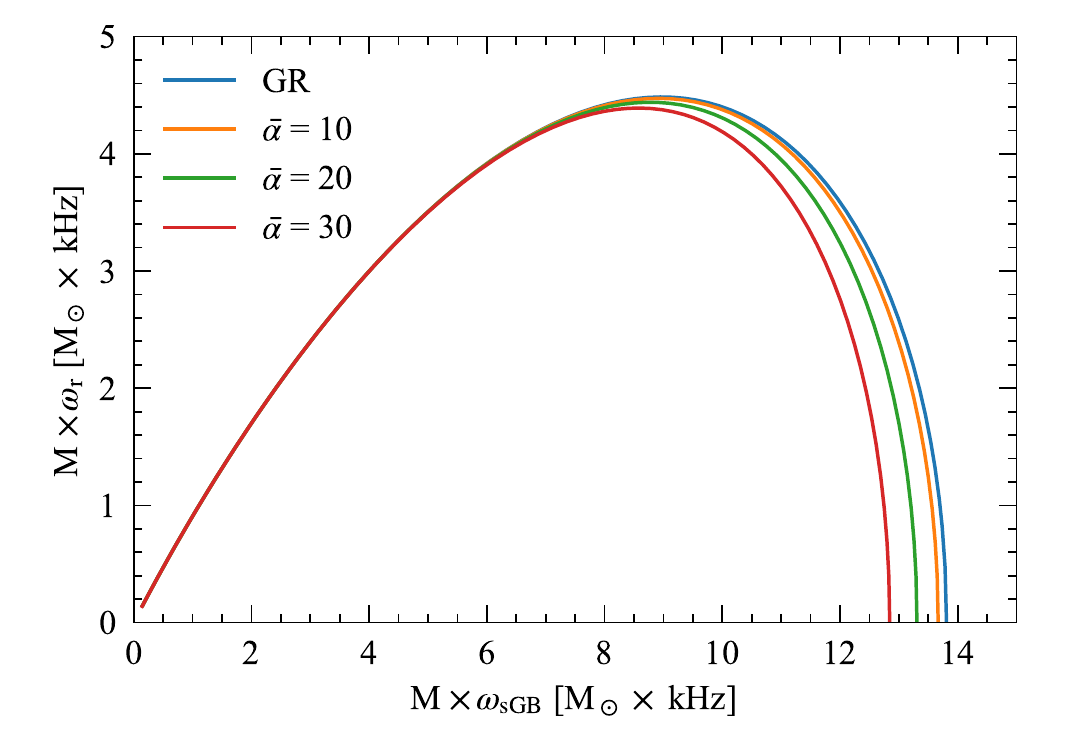}
	\caption{Orbital frequencies $\Omega_r$ versus
    $\Omega_{\rm sGB}$ for a NS of mass $1.4\,M_{\odot}$ to leading
    order in $\zeta$.}
	\label{fig:QPO}
\end{figure}

In addition to these two frequencies, there is often a third one that is important in QPOs
and measures the rate of periastron precession of the orbit.  This precession frequency can be
found via
\begin{equation}
	\Omega_{\rm per} = \Omega_{\rm sGB} - \Omega_r\,,
\end{equation}
and it is usually important in lower frequency QPOs\footnote{Some models treat this frequency as stemming
from inhomogeneities near the inner accretion disk boundary, causing a beat frequency~\cite{Lamb}.
However, this was found to be inconsistent with observations~\cite{Mendez:1998pd}.}~\cite{Glampedakis:2016pes}.
With these three frequencies in hand, one could imagine using the observation of QPOs
to place constraints on sGB.  Figure~\ref{fig:QPO} depicts two of our
frequencies against one another (in dimensionless units) and illustrates how
there are noticeable deviations from the GR predictions as $\alpha$
increases. Observe that the frequencies approach each other when either of them is small,
since here one approaches the weak-field regime described in Eq.~\eqref{eq:KeplerApprox}.

One may present these frequencies in terms of an observable quantity, namely the
dimensionless linear orbital velocity $v$, as done
in~\cite{Ryan:1995wh,Glampedakis:2016pes}.
By introducing the orbital velocity as $v=\left(M\Omega_{\rm sGB}\right)^{1/3}$,
we may reexpress the ratio of the precession frequency to the orbital frequency as
a series in velocity to obtain
\begin{align}
	\frac{\Omega_{\rm per}}{\Omega_{\rm sGB}} &= 3\,v^2 + \frac{9}{2}\,v^4 + \frac{27}{2}\,v^6 + \frac{405}{8}\,v^8 + \frac{1701}{8}\,v^{10} \nonumber \\
	&\quad  + \left(\frac{15309}{16} + 384\,\zeta\right)v^{12} + \order{v^{14}}\,,
\end{align}
where the modification to the GR solution again is suppressed by a high power of velocity
that is consistent with the expansion of Eq.~\eqref{eq:KeplerApprox}. As before, the largest
deviations will then occur for observables that are sensitive to physics near the surface of the NS,
i.e.~where the orbital velocity is not extremely small.

\subsection{Light bending}
\begin{figure}[t]
	\centering
	\includegraphics[width=0.35\textwidth]{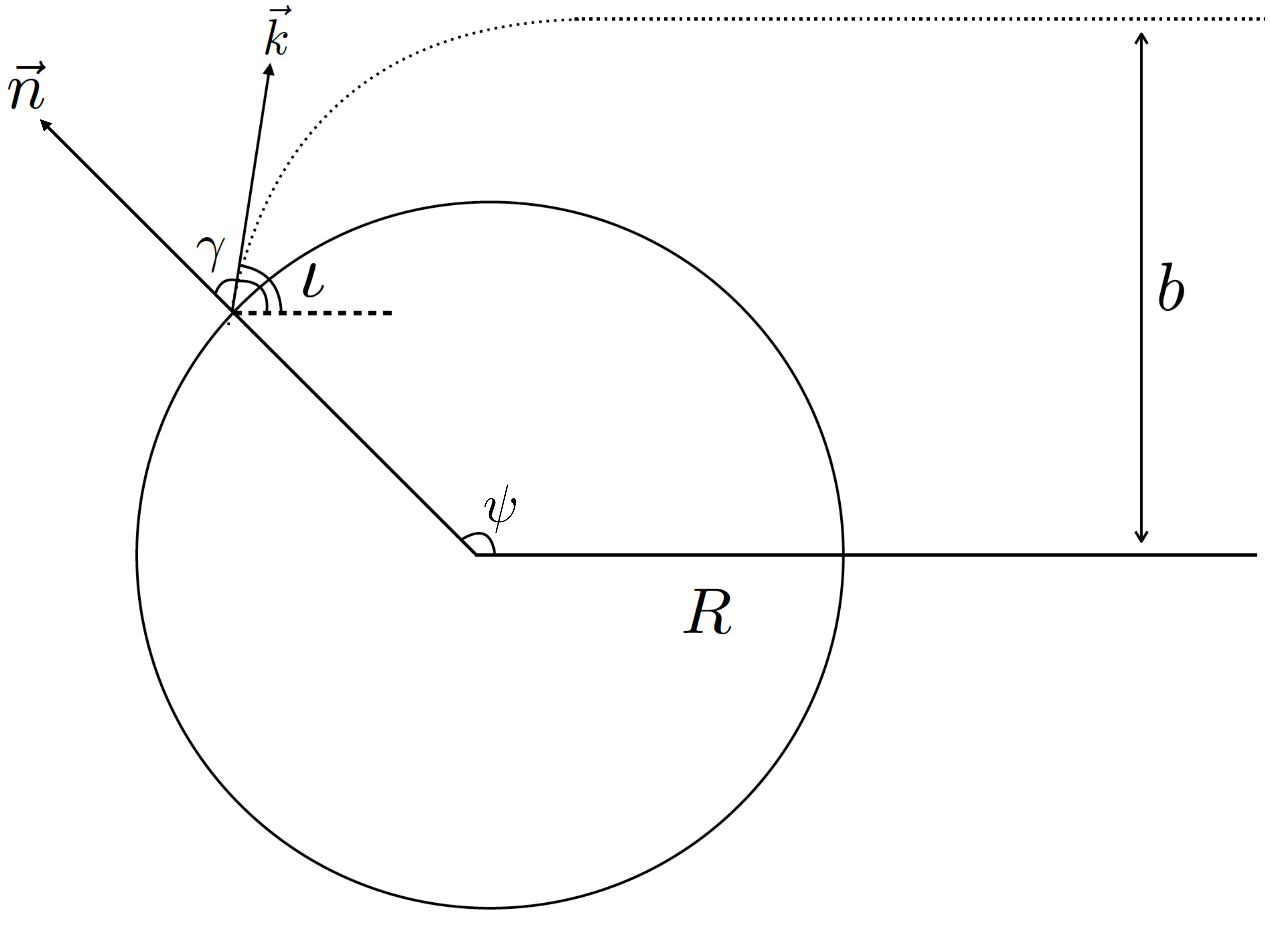}
    \caption{Diagram of emitted photon trajectory.  A photon emitted in the
    direction $\vec{k}$ from the surface of the star orthogonal to $\vec{n}$
    will have its trajectory bent by an angle $\iota = \psi - \gamma$ to an impact
    parameter of $b$.}
	\label{fig:PhotonTrajectory}
\end{figure}

Let us now consider photon motion in the sGB exterior spacetime, as depicted in Fig.~\ref{fig:PhotonTrajectory}.
Imagine then a photon leaving the surface of the NS along the unit vector $\vec{k}$, which makes an angle $\gamma$ with
the unit vector $\vec{n}$ normal to the star's surface. The angle $\psi$, between $\vec{n}$ and the
line of sight, is an important quantity in astrophysical applications. For instance,
when $\gamma=\pi/2$, $\psi=\psi_{\rm crit}$ is the critical angle between the
line of sight and the normal to the surface beyond which the photon cannot reach the observer.
This allows one to define a visible fraction of the star as
\begin{equation}
\label{eq:CritFraction}
	\varsigma \equiv \frac{1}{2}\left[1-\cos\left(\psi_{\rm crit}\right)\right]\,.
\end{equation}
Moreover, in the context of pulse profile modeling, photons emitted by the
hot spot can only reach the observer when the are emitted if emitted with
$\cos \psi > \cos \psi_{\rm crit}$~\cite{Beloborodov:2002mr}.

Let us now derive an expression for $\psi$. We again restrict attention to equatorial orbits,
such that $\theta=0$ and $\dot{\theta}=0$, and change notation $\phi \to \psi$ in
Eqs.~\eqref{eq:conserved_quants} and~\eqref{eq:rDotEqn} with $\epsilon=0$.
Solving for the fraction $\dd \psi/ \dd r$ yields
\begin{equation}
\label{eq:dphidr1}
	\frac{\dd\psi}{\dd r} = \frac{1}{g_{\psi \psi}} \left[-\frac{1}{g_{rr}}\left(\frac{E^2}{L^2}\,\frac{1}{g_{tt}} + \frac{1}{g_{\psi \psi}}\right)\right]^{-1/2}\,.
\end{equation}
Since $E$ and $L$ are constant, we can simplify the above expression through
the emission angle $\gamma$, defined via~\cite{Beloborodov:2002mr}
\begin{equation}
	\label{eq:EmissionAngle}
	\tan^2\left(\gamma\right) = \frac{u^\psi u_\psi}{u^r u_r} \,.
\end{equation}
The above expression allows us to find a relation between $E$, $L$,
and $\gamma$, namely\footnote{The ratio of $L/E$ is also called the
impact parameter~\cite{Sotani:2017rrt}, which is denoted as $b$
in Fig.~\ref{fig:PhotonTrajectory}.}
\begin{equation}
\label{eq:LonE}
	\frac{L}{E} = \sqrt{-\frac{g_{\psi \psi}(R)}{g_{tt}(R)}} \sin\left(\gamma\right)\,,
\end{equation}
where we evaluate the metric functions at the stellar surface. Substituting Eq.~\eqref{eq:LonE}
into Eq.~\eqref{eq:dphidr1} gives a direct relation between $\psi$ and $\gamma$ for a given $R$,
which can be solved to obtain
\begin{equation}
\label{eq:PsiIntegral}
	\psi(R,\gamma) = \int_R^\infty \frac{\dd r}{g_{\psi \psi}}\left[- \frac{1}{g_{rr}} \left(\frac{1}{g_{\psi \psi}}-\frac{g_{\psi \psi}(R) \csc^2\left(\gamma\right)}{g_{tt}(R) \, g_{tt}}\right)\right]\,.
\end{equation}
The integral in Eq.~\eqref{eq:PsiIntegral} may not be straightforward to solve,
even numerically, but following~\cite{Lo:2013ava,Salmi:2018gsn}, we can rewrite it in terms
of the compactness $M/R$ and a new variable $x = \sqrt{1-R/r}$ to ease
the numerical integration.

\begin{figure}[t]
	\centering
	\includegraphics[width=\columnwidth]{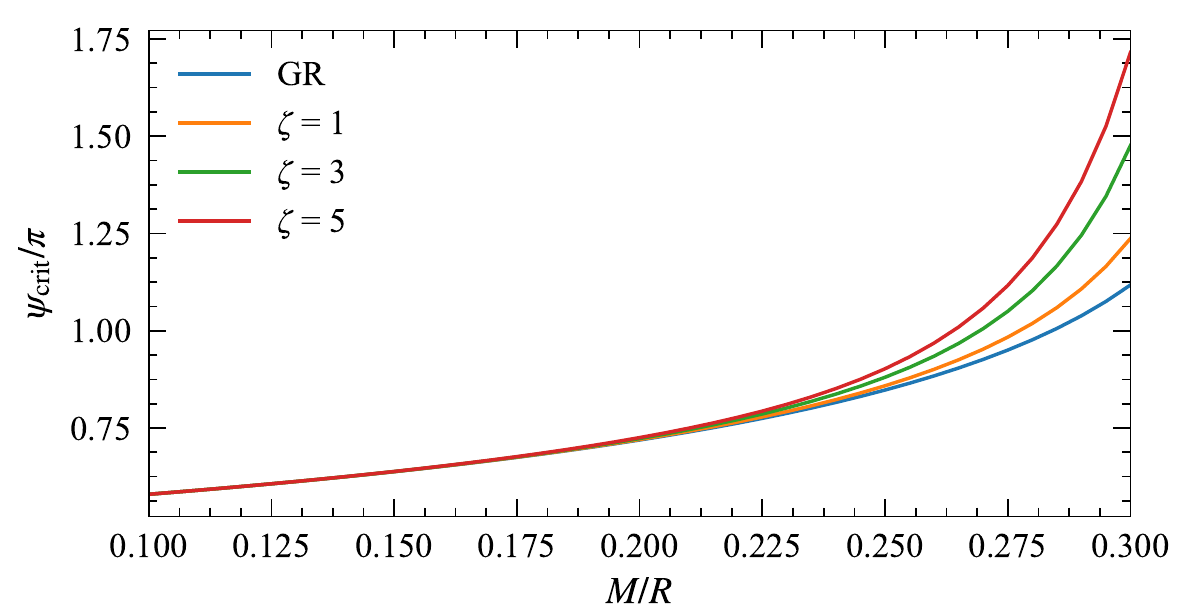}
	\caption{Critical angle as a function of compactness for
    several coupling values $\zeta$.  If we were to fix $\alpha$ as done in
    previous figures, we would also need to specify the NS mass. Since we can
    only measure $M/R$ directly (and not the NS mass) with light bending tests,
    it makes more sense here to fix $\zeta$ instead.}
	\label{fig:CritAngleFig}
\end{figure}

The results of evaluating Eq.~\eqref{eq:PsiIntegral} as a function of the
compactness are shown in Fig.~\ref{fig:CritAngleFig}.
Observe that there is a greater deflection of light for NSs of greater compactness.
This is apparent even in the GR limit, and it is due to the effects of curvature near
compact objects.
However, this effect is enhanced in sGB gravity, increasing with larger
$\zeta$\footnote{The relation between $\zeta$ and $\alpha$ depends on the mass of the NS, which is not specified here.
As a reference, for a $1.4\,M_\odot$ NS, $\bar{\alpha} = \left(10,20,30\right)$ corresponds to $\zeta \approx \left(0.5,2.1,4.6\right)$.},
which dictates how strongly the $\RGB$ correction contributes to the system.
For stars with smaller masses and larger radii, there is a negligible change in
the deflection of light, regardless of the strength of $\zeta$. The curvature of spacetime
near the surface of these NS is simply not large enough even with the quadratic curvature
nature of our theory to cause any deviations that may be detectable in future observations.

\begin{figure}[t]
	\centering
	\includegraphics[width=\columnwidth]{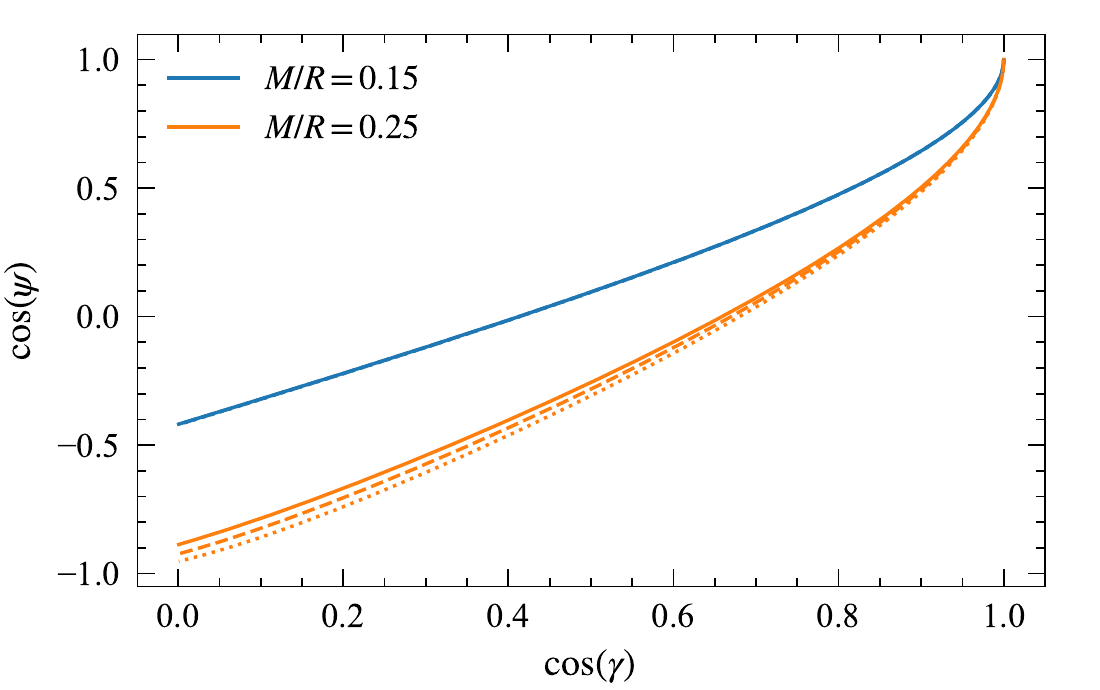}
	\caption{Light bending in sGB gravity.  The solid lines represent
    the GR solution, while the dashed (dotted) lines correspond to $\zeta=2.5$ ($\zeta=5$).
    Deviations from GR are more noticeable when the compactnesses is large and $\zeta$ increases light bending, at fixed emission angle $\gamma$.}
	\label{fig:SchwCompFig}
\end{figure}

We may also look at how light bending in sGB gravity compares to light bending in GR,
as shown in Fig.~\ref{fig:SchwCompFig} for various choices of $\zeta$ values and
two fixed compactnesses. As with Fig.~\ref{fig:CritAngleFig}, there are only tiny deviations
when the compactness is small.  However, NSs with larger compactnesses do present sGB corrections
to light bending that make it stronger relative to GR.

\begin{figure}[t]
	\centering
	\includegraphics[width=\columnwidth]{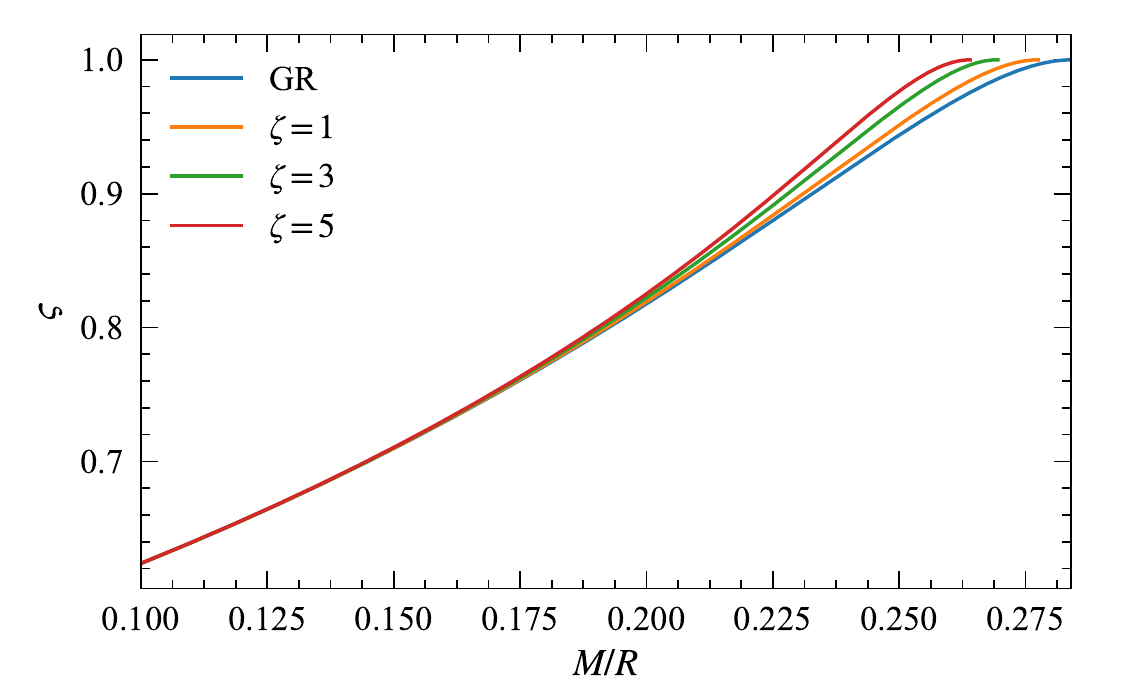}
	\caption{Visible fraction of a star as a function of compactness
    for various coupling strengths.  The lines terminate at the value of compactness
    for which the whole surface of the star becomes visible. Large values of $\zeta$
    require smaller values of compactness for this to happen.}
	\label{fig:VisFracFig}
\end{figure}

As a final calculation, we can also find the visible fraction of the NS surface,
given in Eq.~\eqref{eq:CritFraction}. This is shown in Fig.~\ref{fig:VisFracFig}.
In agreement with our previous results, there is little to be learned about sGB gravity
from observations of low compactness stars. However, as the compactness increases,
so does the effects of the coupling with the Gauss-Bonnet invariant. Likewise, larger values
of the coupling constant lead to larger changes in the visible fraction.
In GR, it is known that for NSs with $M/R \approx 0.28$, strong gravitational
light bending can make the whole surface of the star visible~\cite{1983ApJ...274..846P}.
The effect of the scalar-Gauss-Bonnet coupling is to reduce the necessary compactness
the whole surface of the star to become visible. For instance, when $\zeta = 5$,
this compactness is 0.264.

\section{Conclusions and outlook}
\label{sec:conclusions}

In this paper, we obtained an analytical metric that represents the exterior
spacetime of a NS in sGB gravity as well as an analytical expression for the
scalar field. The metric was derived through a small-coupling perturbative scheme and
depends only on the mass of the NS in question and
the desired strength of the coupling constant. Our metric is valid to $\order{\bar\alpha^{2}}$
and we have outlined how higher-order corrections can be obtained.
We applied the new spacetime to a sample of astrophysical applications, including
the motion of test particle (which is important for instance to model QPOs) and
light bending (which is important to model x-ray pulse profiles generate by hot spots
at the surface of rotating NSs).

Our work opens the door for number a future studies with NSs in sGB
gravity, with the convenience of now being able to treat the metric analytically.
One application could be the development of an effective-one-body (EOB) formalism,
to model NS binaries in sGB, along the lines of the recent work in scalar-tensor
theories~\cite{Julie:2017ucp}.
Having the theory expressed in an EOB framework
allows one to understand the dynamics of the two-body system, the
radiation-reaction components of the system, and knowledge of the
gravitational-waveform emitted from a coalescing binary~\cite{Damour:2012mv}.

Another possible use for the exterior metric is in the modeling of x-ray pulse
profiles as a possible test bed for sGB gravity. These pulse profiles are
generated by the x-ray emission from hot spots on the surface of rotating NS~\cite{1983ApJ...274..846P}
(see~\cite{Poutanen:2008pg,Ozel:2012wu,Watts:2016uzu} for reviews).
As the photons propagate from the surface towards the observer, they probe the
spacetime around the NS which, in principle, can leave detectable deviations in
the observed pulse profile relative to what is predicted in GR.
This possibility was recently explored in the context of scalar-tensor theories~\cite{Sotani:2017bho,Silva:2018yxz,Silva:2019leq}.
In particular, Ref.~\cite{Silva:2019leq} showed that in principle observations made by NICER
can constrain these theories. It would be interesting to see if the
same is true in sGB gravity.

A final observation of interest is the absence of any sGB gravity integration constants in the
final expression for the exterior metric presented in Eqs.~\eqref{eq:FinalMetric_gtt}--\eqref{eq:FinalMetric_grr}.
This is rather unexpected because in other theories (such as in scalar
tensor theories) the exterior metric does depend on charges that must be computed numerically.
In sGB gravity, however, the exterior metric is fully determined in terms of the mass of the star
$M$ and the coupling constant of the theory $\alpha$. A deeper physical or mathematical
understanding of why this is the case in sGB gravity would be most interesting and will be studied elsewhere.

\acknowledgments
This work was supported by NSF Grant No. PHY-1250636 and PHY-1759615, as well as
NASA grants NNX16AB98G and 80NSSC17M0041.
We thank Alejandro C\'ardenas-Avenda\~no, Paolo Pani, Thomas Sotiriou, and
Kent Yagi for helpful discussions.
Computational efforts were performed on the Hyalite High Performance
Computing System, operated and supported by  University Information Technology
Research Cyberinfrastructure at Montana State University.

\bibliographystyle{apsrev4-1}
\bibliography{biblio}

\end{document}